\documentclass{elsart} \textheight= 235mm \textwidth=160mm
\usepackage{graphicx}
\usepackage{amssymb}

\newcommand{\RR}{\mbox{\boldmath$R$}}
\newcommand{\cc}{\mbox{\boldmath$c$}}
\newcommand{\ee}{\mbox{\boldmath$e$}}
\newcommand{\ff}{\mbox{\boldmath$f$}}
\newcommand{\hh}{\mbox{\boldmath$h$}}
\newcommand{\bl}{\mbox{\boldmath$l$}}
\newcommand{\yy}{\mbox{\boldmath$y$}}
\newcommand{\xxx}{\mbox{\boldmath$x$}}
\newcommand{\xx}{\mbox{\boldmath$x$}}
\newcommand{\zz}{\mbox{\boldmath$z$}}
\newcommand{\bb}{\mbox{\boldmath$b$}}
\newcommand{\bg}{\mbox{\boldmath$g$}}
\newcommand{\mm}{\mbox{\boldmath$m$}}
\newcommand{\vl}{\mbox{\boldmath$l$}}
\newcommand{\BB}{\mbox{\boldmath$B$}}
\newcommand{\CC}{\mbox{\boldmath$C$}}
\newcommand{\DD}{\mbox{\boldmath$D$}}
\newcommand{\FF}{\mbox{\boldmath$F$}}
\newcommand{\VV}{\mbox{\boldmath$V$}}
\newcommand{\JJ}{\mbox{\boldmath$J$}}
\newcommand{\PP}{\mbox{\boldmath$P$}}
\newcommand{\HH}{\mbox{\boldmath$H$}}
\newcommand{\LL}{\mbox{\boldmath$L$}}

\newcommand{\NN}{\mbox{\boldmath$N$}}
\newcommand{\UU}{\mbox{\boldmath$U$}}
\newcommand{\ZZ}{\mbox{\boldmath$Z$}}

\newcommand{\bZERO}{\mbox{\boldmath$0$}}
\newcommand{\bOmega}{\mbox{\boldmath$\Omega$}}
\newcommand{\bPi}{\mbox{\boldmath$\Pi$}}
\newcommand{\bPsi}{\mbox{\boldmath$\Psi$}}
\newcommand{\bDelta}{\mbox{\boldmath$\Delta$}}
\newcommand{\bgamma}{\mbox{\boldmath$\gamma$}}
\newcommand{\bmu}{\mbox{\boldmath$\mu$}}
\newcommand{\bnabla}{\mbox{\boldmath$\nabla$}}
\newcommand{\balpha}{\mbox{\boldmath$\alpha$}}
\newcommand{\bbeta}{\mbox{\boldmath$\beta$}}

\begin{document}

\begin{frontmatter}

\title{Invariant grids for reaction kinetics}

\author{Alexander N.\ Gorban}
 \ead{agorban@mat.ethz.ch}
\address{Department of Materials, Institute of Polymer Physics\\
Swiss Federal Institute of Technology, CH-8092 Z\"urich, Switzerland
\\ Institute of Computational Modeling RAS, Krasnoyarsk 660036, Russia}

\author{Iliya V.\ Karlin}
 \ead{ikarlin@mat.ethz.ch}

\address{Department of Materials, Institute of Polymer Physics\\
 Swiss Federal Institute of Technology, CH-8092 Z\"urich, Switzerland}

\author{Andrei Yu. Zinovyev}
\ead{zinovyev@ihes.fr}
\address{Institut des Hautes Etudes Scientifiques, \\ Le
Bois-Marie, 35, route de Chartres, F-91440, Bures-sur-Yvette, France}

\begin{abstract}
In this paper, we review the construction of low-dimensional manifolds of reduced description for
equations of chemical kinetics from the standpoint of the method of invariant manifold (MIM). MIM is
based on a formulation of the condition of invariance as an equation, and its solution by Newton
iterations. A grid-based version of MIM is developed. Generalizations to open systems are suggested.
The set of methods covered makes it possible to effectively reduce description in chemical kinetics.

The most essential new element of this paper is the systematic consideration of a discrete analogue
of the slow (stable) positively invariant manifolds for dissipative systems, {\it invariant grids}.
We describe the Newton method  and the relaxation method for the invariant grids construction. The
problem of the grid correction is fully decomposed into the problems of the grid's nodes correction.
The edges between the nodes appears only in the calculation of the tangent spaces. This fact
determines high computational efficiency of the invariant grids method.

\end{abstract}


\end{frontmatter}

\clearpage

\tableofcontents

\section{Introduction}

In this paper, we present  a general method of constructing the reduced description for dissipative
systems of reaction kinetics and a new method of invariant grids. Our approach is based on the method
of invariant manifold which was developed  in end of 1980th - beginning of 1990th
\cite{GK92,GK92a,GK92b}. Its realization for a generic dissipative systems was discussed in
\cite{GK94,GKIOe01}. This method was applied to a set of problems of classical kinetic theory based
on the Boltzmann kinetic equation \cite{GK94,KDN97,KGDN98}. The method of invariant manifold was
successfully applied to a derivation of reduced description for kinetic equations of polymeric
solutions \cite{ZKD00}. It was also been tested on systems of chemical kinetics
\cite{GKZD00,InChLANL}. In order to construct manifolds of a relatively low dimension, grid-based
representations of manifolds become a relevant option. The idea of invariant grids was suggested
recently in \cite{InChLANL}.

The goal of nonequilibrium statistical physics is the understanding of how a system with many degrees
of freedom acquires a description with a few degrees of freedom. This should lead to reliable methods
of extracting the macroscopic description from a detailed microscopic description.

Meanwhile this general problem is still far from the final solution, it is reasonable to study
simplified models, where, on the one hand, a detailed description is accessible to numerics, on the
other hand, analytical methods designed to the solution of problems in real systems can be tested.

In this paper we address the well known class of finite-dimensional systems known from the theory of
reaction kinetics. These are equations governing a complex relaxation in perfectly stirred closed
chemically active mixtures. Dissipative properties of such systems are characterized with a global
convex Lyapunov function $G$ (thermodynamic potential) which implements the second law of
thermodynamics: As the time $t$ tends to infinity, the system reaches the unique equilibrium state
while in the course of the transition the Lyapunov function decreases monotonically.

While the limiting behavior of the dissipative systems just described is certainly very simple, there
are still interesting questions to be asked about. One of these questions is closely related to the
above general problem of nonequilibrium statistical physics. Indeed, evidence of numerical
integration of such systems often demonstrates that the relaxation has a certain geometrical
structure in the phase space. Namely, typical individual trajectories tend to manifolds of lower
dimension, and further proceed to the equilibrium essentially along these manifolds. Thus, such
systems demonstrate a dimensional reduction, and therefore establish a more macroscopic description
after some time since the beginning of the relaxation.

There are two intuitive ideas behind our approach, and we shall now discuss them informally. Objects
to be considered below are manifolds (surfaces) $\bOmega$ in the phase space of the reaction kinetic
system (the phase space is usually a convex polytope in a finite-dimensional real space). The `ideal'
picture of the reduced description we have in mind is as follows: A typical phase trajectory,
${\cc}(t)$, where $t$ is the time, and $\cc$ is an element of the phase space, consists of two
pronounced segments. The first segment connects the beginning of the trajectory, $\cc(0)$, with a
certain point, $\cc(t_1)$, on the manifold $\Omega$ (rigorously speaking, we should think of
$\cc(t_1)$ not on $\bOmega$ but in a small neighborhood of $\bOmega$ but this is inessential for the
ideal picture). The second segment belongs  to $\bOmega$, and connects the point $\cc(t_1)$ with the
equilibrium $\cc^{\rm eq}={\cc}(\infty)$, $\cc^{\rm eq}\in\bOmega$. Thus, the manifolds appearing in
our ideal picture are ``patterns'' formed by the segments of individual trajectories, and the goal of
the reduced description is to ``filter out'' this manifold.

There are two important features behind this ideal picture. The first feature is the {\it invariance}
of the manifold $\bOmega$: Once the individual trajectory has started on  $\bOmega$, it does not
leaves $\bOmega$ anymore. The second feature is the {\it projecting}: The phase points outside
$\bOmega$ will be projected onto $\bOmega$. Furthermore, the dissipativity of the system provides an
{\it additional} information about this ideal picture: Regardless of what happens on  the manifold
$\bOmega$, the function $G$ was decreasing along each individual trajectory before it reached
$\bOmega$. This ideal picture is the guide to extract slow invariant manifolds.

One more point needs a clarification before going any further. Low dimensional invariant manifolds
exist also for systems with a more complicated dynamic behavior, so why to study the  invariant
manifolds of slow motions for a particular class of purely dissipative systems? The answer is in the
following: Most of the physically significant models include non-dissipative components in a  form of
either a  conservative dynamics, or in the form of external forcing or external fluxes. Example of
the first kind is the free flight of particles on top of the dissipation-producing collisions in the
Boltzmann equation. For the second type of example one can think of irreversible reactions among the
suggested stoichiometric mechanism (inverse process are so unprobable that we discard  them
completely thereby effectively ``opening'' the system to the remaining irreversible flux). For all
such systems, the present method is applicable almost without special refinements, and bears the
significance that invariant manifolds are constructed as  a ``deformation'' of the relevant manifolds
of slow motion of the purely dissipative dynamics. Example of this construction for open systems is
presented below in section \ref{open}. Till then we focus on the purely dissipative case for the
reason just clarified.

The most essential new element of this paper is the systematic consideration of a discrete analogue
of the slow (stable) positively invariant manifolds for dissipative systems, {\it invariant grids}.
These invariant grids were introduced in the \cite{InChLANL}. Here we will describe the Newton method
subject to incomplete linearization and the relaxation methods for the invariant grids. It is worth
to mention, that the problem of the grid correction is fully decomposed into the problems of the
grid's nodes correction. The edges between the nodes appears only in the calculation of the tangent
spaces. This fact determines high computational efficiency of the invariant grids method.

Due to the famous Lyapunov auxiliary theorem \cite{Lya,Kazantzis00}  we can construct analytical
invariant manifolds for kinetic equations with analytical right hand side. Moreover, the analycity
can serve as a ``selection rule" for selection the unique analytic positively invariant manifold from
the infinite set of smooth positively invariant manifolds. The analycity gives a possibility to use
the powerful  technique of analytical continuation and Carleman's formulae \cite{Aiz,GR99,GapsGRMD}.
It leads us to {\it superresolution effects}: A small grid may be sufficient to present an ``large"
analytical manifold immersed in the whole space.

The paper is organized as follows. In the section \ref{OUTLINE}, we review the reaction kinetics
(section \ref{KINETICS}), and discuss the main methods of model reduction in chemical kinetics
(section \ref{reduction_review}). In particular, we present two general versions of extending
partially equilibrium manifolds to a single relaxation time model in the whole phase space, and
develop a thermodynamically consistent version of the intrinsic low-dimensional manifold (ILDM)
approach. In the section \ref{MIMG} we  review the method of invariant manifold in the way
appropriate to this class of nonequilibrium systems. In the sections \ref{THERMO} and \ref{DC} we
give some details on the two relatively independent parts of the method, the thermodynamic projector,
and the iterations for solving the invariance equation.

We also describe a general symmetric linearization procedure for the invariance equation, and discuss
its relevance to the picture of decomposition of motions. In the section \ref{MIM}, these two
procedures are combined into an unique algorithm. In the section \ref{EX}, we demonstrate an
illustrative example of  analytic computations for a model catalytic reaction. In the section
\ref{relax} we introduce the relaxation method for solution the invariance equation. This  relaxation
method is an alternative to the Newton iteration method. In the section \ref{PARAMETERIZATION} we
demonstrate how the {\it thermodynamic projector} is constructed without the a priori
parameterization of the manifold\footnote{ This thermodynamic projector is the unique operator which
transforms the arbitrary vector field equipped with the given Lyapunov function into a vector field
with the same Lyapunov function (and also this happens on any manifold which is not tangent to the
level of the Lyapunov function).}. This result is essentially used in the section \ref{grid} where we
introduce a computationally effective grid-based method to construct invariant manifolds. It is the
central section of the paper. We present the Newton method and the relaxation method for the grid
construction. The Carleman formulas for analytical continuation a manifold from a grid are proposed.
Two examples of kinetic equations are analyzed: a two-dimensional catalytic reaction (four species,
two balances) and a four-dimensional oxidation reaction (six species, two balances).

In the section \ref{open} we describe an extension of the method of invariant manifold to open
systems. Finally, results are discussed in the section \ref{conclusion}.

\section{Equations of chemical kinetics and their reduction}
\label{OUTLINE}

\subsection{Outline of the dissipative reaction kinetics}
\label{KINETICS}

We begin with an outline  of the reaction kinetics (for details see e.\ g.\ the book of
\cite{YBGE91}). Let us consider a closed system with $n$ chemical species ${\rm A}_1,\dots,{\rm
A}_n$, participating in a complex reaction. The complex reaction is represented by the following
stoichiometric mechanism:
\begin{equation}
\label{stoi} \alpha_{s1}{\rm A}_1+\ldots+\alpha_{sn}{\rm A}_n\rightleftharpoons \beta_{s1}{\rm
A}_1+\ldots+\beta_{sn}{\rm A}_n,
\end{equation}
where the index $s=1,\dots,r$ enumerates the reaction steps, and where integers, $\alpha_{si}$ and
$\beta_{si}$,  are stoichiometric coefficients. For each reaction step $s$, we introduce
$n$--component vectors $\balpha_s$ and $\bbeta_s$ with components $\alpha_{si}$ and $\beta_{si}$.
Notation $\mbox{\boldmath$\gamma$}_s$  stands for the vector with integer components
$\gamma_{si}=\beta_{si}-\alpha_{si}$ (the stoichiometric vector). We adopt an abbreviated notation
for the standard scalar product of the $n$-component vectors:
\[(\xx,\yy)=\sum_{{i=1}}^{n}x_iy_i.\]

The system is described by the  $n$-component concentration vector $\cc$, where the component
$c_i\ge0$ represents the concentration of the specie ${\rm A}_i$. Conservation laws impose linear
constraints on admissible vectors $\cc$ (balances):
\begin{equation}
\label{conser} (\bb_i, \cc)=B_i,\ i=1,\dots,l,
\end{equation}
where $\bb_i$ are fixed and linearly independent vectors, and $B_i$ are given scalars. Let us denote
as $\BB$ the set of vectors which satisfy the conservation laws (\ref{conser}):
\[
\BB=\left\{\cc|(\bb_1,\cc)=B_1,\dots, (\bb_l,\cc)=B_l\right\}.
\]

The phase space $\VV$ of the system is the intersection of the cone of $n$-dimensional vectors with
nonnegative components, with the set $\BB$, and ${\rm dim}\VV=d=n-l$. In the sequel, we term a vector
$\cc\in\VV$ the state of the system. In addition, we assume that each of the conservation laws is
supported by each elementary reaction step, that is
\begin{equation}
\label{sep} (\bgamma_s,\bb_i)=0,
\end{equation}
for each pair of vectors $\bgamma_s$ and $\bb_i$.

Reaction kinetic equations describe variations of the states in time. Given the stoichiometric
mechanism (\ref{stoi}), the reaction kinetic equations read:
\begin{equation}
\label{reaction} \dot{\cc}=\JJ(\cc),\ \JJ(\cc)=\sum_{s=1}^{r}\bgamma_sW_s(\cc),
\end{equation}
where dot denotes the time derivative, and $W_s$  is the reaction rate function of the step $s$. In
particular, the mass action law suggests the polynomial form of the reaction rates:
\begin{equation}
\label{MAL} W_s=k^+_s \prod_{i=1}^{n}c_i^{\alpha_i} - k^-_s \prod_{i=1}^{n} c_i^{\beta_i},
\end{equation}
where $k^+_s$ and $k^-_s$ are the constants of the direct and of the inverse reactions rates of the
$s$th reaction step. The phase space $\VV$ is positive-invariant of the system (\ref{reaction}): If
$\cc(0)\in\VV$, then $\cc(t)\in\VV$ for all the times $t>0$.

In the sequel, we assume that the kinetic equation (\ref{reaction}) describes  evolution towards the
unique equilibrium state, $\cc^{\rm eq}$, in the interior of the phase space $\VV$. Furthermore, we
assume that there exists a strictly convex function $G(\cc)$ which decreases monotonically in time
due to Eq.\ (\ref{reaction})\footnote{With some abuse of language, we can term the functional $-G$
the entropy, although it is a different functional for non-isolated systems. We recall that
thermodynamic Lyapunov functions are well defined not just for isolated systems. Such functionals are
easily constructed also for systems which exchange energy and/or matter with a larger equilibrium
system (with a thermostat, for example). In such a case, the thermodynamic Lyapunov function is
constructed as the entropy of the minimal closed system containing the system under consideration
\cite{G84}. In particular, the free energy and the free enthalpy (the Gibbs and the Helmholz
energies, respectively) can be constructed in this manner. They are are identical with the entropy of
the minimal closed system containing the given system within the accuracy of multiplication with a
factor which remains constant in time, and subtracting a constant.} :
\begin{equation}
\label{Htheorem} \dot{G}= (\bnabla G(\cc),\JJ(\cc)) \leq 0,
\end{equation}
Here $\bnabla G$ is the vector of partial derivatives $\partial G/\partial c_i$, and the convexity
assumes that the $n\times n$ matrices
\begin{equation}
\label{MATRIX} \HH_{\cc}=\|\partial^2G(\cc)/\partial c_i\partial c_j\|,
\end{equation}
are positive definite for all $\cc\in\VV$. In addition, we assume that the matrices (\ref{MATRIX})
are invertible if $\cc$ is taken in the interior of the phase space.

The function $G$ is the Lyapunov function of the system (\ref{reaction}), and $\cc^{\rm eq}$ is the
point of global minimum of the function $G$ in the phase space $\VV$. Otherwise stated, the manifold
of equilibrium states $\cc^{\rm eq}(B_1,\dots,B_l)$ is the solution to the variational problem,
\begin{equation}
\label{EQUILIBRIUM} G\to{\rm min}\ {\rm for\ }(\bb_i,\cc)=B_i,\ i=1,\dots,l.
\end{equation}
For each fixed value of the conserved quantities $B_i$, the solution is unique. In many cases,
however, it is convenient to consider the whole equilibrium manifold, keeping the conserved
quantities as parameters.

For example, for perfect systems in a constant volume under a constant temperature, the Lyapunov
function $G$ reads:
\begin{equation}
\label{gfun} G=\sum_{i=1}^{n}c_i[\ln(c_i/c^{\rm eq}_i)-1].
\end{equation}

It is important to stress that $\cc^{\rm eq}$ in Eq.\ (\ref{gfun})
 is an {\it arbitrary} equilibrium of the system,
under arbitrary values of the balances. In order to compute $G(\cc)$, it is unnecessary to calculate
the specific equilibrium $\cc^{\rm eq}$ which corresponds to the initial state $\cc$. Moreover, for
ideal systems, function $G$ is constructed from the thermodynamic data of individual species, and, as
the result of this construction, it turns out that it has the form of Eq.\ (\ref{gfun}). Let us
mention here the classical formula for the free energy $F=RTVG$:
\begin{equation}
F=VRT\sum_{i=1}^{n}c_i[(\ln(c_iV_{{\rm Q}\ i})-1)+F_{{\rm int}\ i}(T)],
\end{equation}
where $V$ is the volume of the system, $T$ is the temperature, $V_{{\rm Q}\ i}=
N_0(2\pi\hbar^2/m_ikT)^{3/2}$ is the quantum volume of one mole of the specie $A_i$, $N_0$ is the
Avogadro number, $m_i$ is the mass of the molecule of ${\rm A}_i$, $R=kN_0$, and $F_{{\rm int}\
i}(T)$ is the free energy of the internal degrees of freedom per mole of ${\rm A}_i$.

Finally, we recall an important generalization of the mass action law (\ref{MAL}), known as the
Marcelin-De Donder kinetic function. This generalization was developed in \cite{Feinberg72} based on
ideas of the thermodynamic theory of affinity \cite{DeDonder36}. We use the kinetic function
suggested in \cite{BGY82}. Within this approach, the functions $W_s$ are constructed as follows: For
a given strictly convex function $G$, and for a given stoichiometric mechanism (\ref{stoi}), we
define the gain ($+$) and the loss ($-$) rates of the $s$th step,
\begin{equation}
\label{MDD} W_s^{+}=\varphi_s^+ \exp[( \nabla G,\mbox{\boldmath$\alpha$}_s)],\quad
W_s^{-}=\varphi_s^-\exp[( \nabla G,\mbox{\boldmath$\beta$}_s)],
\end{equation}
where $\varphi_s^{\pm}>0$ are kinetic factors. The Marcelin-De Donder kinetic function reads:
$W_s=W_s^+-W_s^-$, and the right hand side of the kinetic equation (\ref{reaction}) becomes,
\begin{equation}
\label{KINETIC MDD} \JJ=\sum_{s=1}^{r}\bgamma_s \{\varphi_s^+ \exp[(\bnabla G,\balpha_s)]\!-
\varphi_s^-\exp[(\bnabla G,\bbeta_s)]\}.
\end{equation}
For the Marcelin-De Donder reaction rate (\ref{MDD}), the dissipation inequality (\ref{Htheorem})
reads:
\begin{equation}
\label{HMDD} \dot{G}=\sum_{s=1}^{r} [(\bnabla G,\bbeta_s) - (\bnabla G,\balpha_s)] \left\{\varphi_s^+
e^{(\bnabla G,\balpha_s)}- \varphi_s^-e^{(\bnabla G,\bbeta_s)}\right\}\le 0.
\end{equation}
The kinetic factors $\varphi_s^{\pm}$ should satisfy certain conditions in order to make valid the
dissipation inequality (\ref{HMDD}). A well known sufficient condition is the detail balance:
\begin{equation}
\label{DB} \varphi_s^+=\varphi_s^-,
\end{equation}
other sufficient conditions are discussed in detail elsewhere \cite{YBGE91,G84,K89,K93}. For the
function  $G$ of the form (\ref{gfun}), the Marcelin-De Donder equation casts into the more familiar
mass action law form (\ref{MAL}).

\subsection{The problem of reduced description in chemical kinetics}
\label{reduction_review}

What does it mean, ``to reduce the description of a chemical system''? This means the following:
\begin{enumerate}
\item To shorten  the list of species.
This, in turn, can be achieved in two ways:

(i) To eliminate inessential components from the list;

(ii) To lump some of the species into integrated components.

\item To shorten the list of reactions. This also can be done in several ways:

(i) To eliminate inessential reactions, those which do not significantly influence the reaction
process;

(ii) To assume that some of the reactions ``have been already completed'', and that the equilibrium
has been reached along their paths (this leads to dimensional reduction because the rate constants of
the ``completed'' reactions are not used thereafter, what one needs are equilibrium constants only).

\item To decompose the motions into fast and slow, into independent (almost-independent)
and slaved etc. As the result of such a decomposition, the system admits a study ``in parts''. After
that, results of this study are combined into a joint picture. There are several approaches which
fall into this category: The famous method of the quasi-steady state (QSS), pioneered by Bodenstein
and Semenov and explored in considerable detail  by many authors, in particular, in
\cite{Bowen63,Chen88,Segel89,Fraser88,Roussel90}, and many others; the quasi-equilibrium
approximation \cite{Orlov84,G84,Volpert85,Fraser88,K89,K93}; methods of sensitivity analysis
\cite{Rabitz83,Lam94}; methods based on the derivation of the so-called intrinsic low-dimensional
manifolds (ILDM, as suggested in \cite{Maas92}). Our method of invariant manifold (MIM,
\cite{GK92,GK92a,GK92b,GK94,GKZD00,GKIOe01}) also belongs to this kind of methods.

\end{enumerate}
Why to reduce description in the times of supercomputers?

First, in order to gain understanding. In the process of reducing the description one is often able
to extract the essential, and the mechanisms of the processes under study become more transparent.
Second, if one is given the detailed description of the system, then one should be able also to solve
the initial-value problem for this system. But what should one do in the case where the the system is
representing just a point in a three-dimensional flow? The problem of reduction becomes particularly
important for modeling the spatially distributed physical and chemical processes. Third, without
reducing the kinetic model, it is impossible to construct this model. This statement seems paradoxal
only at the first glance: How come, the model is first simplified, and is constructed only after the
simplification is done? However, in practice, the typical for a mathematician statement of the
problem, (Let the system of differential equations be {\it given}, then ...) is rather rarely
applicable in the chemical engineering science for detailed kinetics. Some reactions are known
precisely, some other - only hypothetically. Some intermediate species are well studied, some others
- not, it is not known much about them. Situation is even worse with the reaction rates. Quite on the
contrary, the thermodynamic data (energies, enthalpies, entropies, chemical potentials etc) for
sufficiently rarefied systems are quite reliable. Final identification of the model is always done on
the basis of comparison with the experiment and with a help of fitting. For this purpose, it is
extremely important to reduce the dimension of the system, and to reduce the number of tunable
parameters. The normal logics of modeling for the purpose of chemical engineering science is the
following: Exceedingly detailed but coarse with respect to parameters system $\to$ reduction $\to$
fitting $\to$ reduced model with specified parameters (cycles  are allowed in this scheme, with
returns from fitting to more detailed models etc). A  more radical viewpoint is also possible: In the
chemical engineering science, detailed kinetics is impossible, useless, and it does not exist. For a
recently published discussion on this topic see \cite{Levenspiel99,Levenspiel00}; \cite{Y00}.

Alas, with a mathematical statement of the problem
 related to reduction, we  all have to begin with the usual:
Let the system of differential equations be given ... . Enormous difficulties related to the question
of how well the original system is modeling the real kinetics remain out of focus of these studies.

Our present work is devoted to studying reductions in a given system of kinetic equations to
invariant manifolds of slow motions. We begin with a brief discussion of existing approaches.

\subsection{Partial equilibrium approximations}\label{partial_eq}

{\it Quasi-equilibrium with respect to reactions} is constructed as follows: From the list of
reactions (\ref{stoi}), one selects those which are assumed to equilibrate first. Let they be indexed
with the numbers $s_1,\dots,s_k$. The quasi-equilibrium manifold is defined by the system of
equations,
\begin{equation}
\label{st1} W^+_{s_i}=W^-_{s_i},\ i=1,\dots,k.
\end{equation}
This system of equations looks particularly elegant when written in terms of conjugated (dual)
variables,  $\bmu=\bnabla G$:
\begin{equation}
\label{st2} ( \bgamma_{s_i},\bmu )=0,\ i=1,\dots,k.
\end{equation}
In terms of conjugated variables, the quasi-equilibrium manifold forms a linear subspace. This
subspace, $L^{\perp}$, is the orthogonal completement to the linear envelope of vectors, $L={\rm
lin}\{\bgamma_{s_1},\dots,\bgamma_{s_k}\}$.

{\it Quasi-equilibrium with respect to species} is constructed practically  in the same way but
without selecting the subset of reactions. For a given set of species, $A_{i_1}, \dots, A_{i_k}$, one
assumes that they evolve fast to equilibrium, and remain there. Formally, this means that in the
$k$-dimensional subspace of the space of concentrations with the coordinates $c_{i_1},\dots,c_{i_k}$,
one constructs the subspace $L$ which is defined by the balance equations, $( \bb_i,\cc)=0$. In terms
of the conjugated variables, the quasi-equilibrium manifold, $L^{\perp}$, is defined by equations,
\begin{equation}
\label{qe1} \bmu\in L^{\perp},\ (\bmu=(\mu_1,\dots,\mu_n)).
\end{equation}
The same quasi-equilibrium manifold can be also defined with the help of fictitious reactions: Let
$\bg_1,\dots,\bg_q$ be a basis in $L$. Then Eq.\ (\ref{qe1}) may be rewritten as follows:
\begin{equation}
\label{qe2} ( \bg_i,\bmu )=0,\ i=1,\dots,q.
\end{equation}

{\it Illustration:} Quasi-equilibrium with respect to reactions in hydrogen oxidation: Let us assume
equilibrium with respect to dissociation reactions, ${\rm H}_2\rightleftharpoons 2{\rm H}$, and,
${\rm O}_2\rightleftharpoons 2{\rm O}$, in some subdomain of reaction conditions. This gives:
\[k_1^+c_{{\rm H}_2}=k_1^-c^2_{\rm H},\ k_2^+c_{{\rm O}_2}=k_2^-c_{\rm O}^2.\]
Quasi-equilibrium with respect to species: For the same reaction, let us assume equilibrium over
${\rm H}$, ${\rm O}$, ${\rm OH}$, and ${\rm H}_2{\rm O}_2$, in a subdomain of reaction conditions.
Subspace $L$ is defined by balance constraints:
\[ c_{\rm H}+c_{\rm OH}+2c_{{\rm H}_2{\rm O}_2}=0,\ c_{\rm O}+c_{\rm OH}
+2c_{{\rm H}_2{\rm O}_2}=0.\] Subspace $L$ is two-dimensional. Its basis, $\{\bg_1,\bg_2\}$ in the
coordinates $c_{\rm H}$, $c_{\rm O}$, $c_{\rm OH}$, and $c_{{\rm H}_2{\rm O}_2}$ reads:
\[
\bg_1=(1,1,-1,0),\quad \bg_2=(2,2,0,-1).
\]
Corresponding Eq.\ (\ref{qe2}) is:
\[ \mu_{\rm H}+\mu_{\rm O}=\mu_{\rm OH},\ 2\mu_{\rm H}+2\mu_{\rm O}=
\mu_{{\rm H}_2{\rm O}_2}.\]

{\it General construction of the quasi-equilibrium manifold}: In the space of concentration, one
defines a subspace $L$ which satisfies the balance constraints:
\[ ( \bb_i,L)\equiv0.\]
The orthogonal complement of $L$ in the space with coordinates $\bmu=\bnabla G$ defines then the
quasi-equilibrium manifold $\bOmega_{L}$. For the actual computations, one requires the inversion
from $\bmu$ to $\cc$. Duality structure $\bmu\leftrightarrow\cc$ is well studied by many authors
\cite{Orlov84,DKN97}.

{\it Quasi-equilibrium projector.} It is not sufficient to just derive the manifold, it is also
required to define a {\it projector} which would transform the vector field defined on the space of
concentrations to a vector field on the manifold. Quasi-equilibrium manifold consists of points which
minimize $G$ on the affine spaces of the form $\cc+L$. These affine  planes are hypothetic planes of
fast motions ($G$ is decreasing in the course of the fast motions). Therefore, the quasi-equilibrium
projector maps the whole space of concentrations on $\bOmega_L$ parallel to $L$. The vector field is
also projected onto the tangent space of $\bOmega_L$ parallel to $L$.

Thus, the quasi-equilibrium approximation implies the decomposition of motions into the fast -
parallel to $L$, and the slow - along the quasi-equilibrium manifold. In order to construct the
quasi-equilibrium approximation, knowledge of reaction rate constants of ``fast'' reactions is not
required (stoichiometric vectors of all these fast reaction are in $L$, $\bgamma_{{\rm fast}}\in L$,
thus, knowledge of $L$ suffices), one only needs some confidence in that they all are sufficiently
fast \cite{Volpert85}. The quasi-equilibrium manifold itself is constructed based on the knowledge of
$L$ and of $G$. Dynamics on the quasi-equilibrium manifold is defined as the quasi-equilibrium
projection of the ``slow component'' of kinetic equations (\ref{reaction}).

\subsection{Model equations}

The assumption behind  the quasi-equilibrium is the hypothesis of the decomposition of motions into
fast and slow. The quasi-equilibrium approximation itself describes slow motions. However, sometimes
it becomes necessary to restore to the whole system, and to take into account the fast motions as
well. With this, it is desirable to keep intact one of the important advantages of the
quasi-equilibrium approximation -
 its independence of the rate constants of fast reactions.
For this purpose, the detailed fast kinetics is replaced by a model equation ({\it single relaxation
time approximation}).

{\it Quasi-equilibrium models} (QEM) are constructed as follows: For each concentration vector $\cc$,
consider the affine manifold, $\cc+L$. Its intersection with the quasi-equilibrium manifold
$\bOmega_L$ consists of one point. This point delivers the  minimum to $G$ on $\cc+L$. Let us denote
this point as $\cc^*_L(\cc)$. The equation of the quasi-equilibrium model reads:
\begin{equation}
\label{QEmodel} \dot{\cc}=-\frac{1}{\tau}[\cc-\cc^*_L(\cc)]+\sum_{{\rm
slow}}\bgamma_{s}W_s(\cc^*_L(\cc)),
\end{equation}
where $\tau>0$ is the relaxation time of the fast subsystem. Rates of slow reactions are computed in
the points $\cc^*_L(\cc)$ (the second term in the right hand side of Eq.\ (\ref{QEmodel}), whereas
the rapid motion is taken into account by a simple relaxational term (the first term in the right
hand side of Eq.\ (\ref{QEmodel}). The most famous model kinetic equation is the BGK equation in the
theory of the Boltzmann equation \cite{BGK}. The general theory of the quasi-equilibrium models,
including proofs of their thermodynamic consistency, was constructed in \cite{GK92c,GK94a}.

{\it Single relaxation time gradient models} (SRTGM) were considered in \cite{AK00,AK02,AK02a} in the
context of the lattice Boltzmann method for hydrodynamics. These models are aimed at improving the
obvious drawback of quasi-equilibrium models (\ref{QEmodel}): In order to construct the QEM, one
needs to compute the function,
\begin{equation}
\label{QEA}
 \cc^*_L(\cc)=\arg\min_{\xxx\in \cc+L,\ \xxx>0}G(\xxx).
\end{equation}
This is a convex programming problem. It does not always has a closed-form solution.

Let $\bg_1,\dots,\bg_k$ is the orthonormal basis of $L$. We denote as $\DD(\cc)$ the  $k\times k$
matrix with the elements $( \bg_i,\HH_{\cc}\bg_j)$, where $\HH_{\cc}$ is the matrix of second
derivatives of $G$ (\ref{MATRIX}). Let $\CC(\cc)$ be the inverse of $\DD(\cc)$. The single relaxation
time gradient model has the form:
\begin{equation}
\dot{\cc}=-\frac{1}{\tau}\sum_{i,j}\bg_i\CC(\cc)_{ij}( \bg_j,\bnabla G) +\sum_{{\rm
slow}}\bgamma_{s}W_s(\cc).\label{SRTGM}
\end{equation}
The first term drives the system to the minimum of $G$ on $\cc+L$, it does not require solving the
problem (\ref{QEA}), and its spectrum in the quasi-equilibrium is the same as in the
quasi-equilibrium model (\ref{QEmodel}). Note that the slow component is evaluated in the ``current''
state $\cc$.

The models (\ref{QEmodel}) and (\ref{SRTGM})
 lift the quasi-equilibrium approximation to a kinetic equation
by approximating the fast dynamics with a single ``reaction rate constant'' - relaxation time $\tau$.

\subsection{Quasi-steady state approximation}\label{QSS}

The quasi-steady state approximation (QSS) is a tool used in a huge amount of works. Let us split the
list of species in two groups: The basic and the intermediate (radicals etc). Concentration vectors
are denoted accordingly, $\cc^{\rm s}$ (slow, basic species), and $\cc^{\rm f}$ (fast, intermediate
species). The concentration vector $\cc$ is the direct sum, $\cc=\cc^{\rm s}\oplus\cc^{\rm f}$. The
fast subsystem is Eq.\ (\ref{reaction}) for the component $\cc^{\rm f}$ at fixed values of $\cc^{{\rm
s}}$. If it happens that this way defined fast subsystem relaxes to a stationary state, $\cc^{\rm
f}\to\cc^{\rm f}_{\rm qss}(\cc^{\rm s})$, then the assumption that $\cc^{\rm f}=\cc^{\rm f}_{\rm
qss}(\cc)$ is precisely  the QSS assumption. The slow subsystem is the part of the system
(\ref{reaction}) for $\cc^{\rm s}$, in the right hand side of which the component $\cc^{\rm f}$ is
replaced with $\cc^{\rm f}_{\rm qss}(\cc)$. Thus, $\JJ=\JJ_{\rm s}\oplus\JJ_{\rm f}$, where
\begin{eqnarray}
\dot{\cc}^{\rm f}&=&\JJ_{\rm f}(\cc^{\rm s}\oplus\cc^{\rm f}), \ \cc^{\rm s}={\rm const}; \quad
\cc^{\rm f}\to\cc^{\rm f}_{\rm qss}(\cc^{\rm s});\label{fast}\\ \dot{\cc}^{\rm s}&=&\JJ_{\rm
s}(\cc^{\rm s}\oplus\cc_{\rm qss}^{\rm f}(\cc^{\rm s})). \label{slow}
\end{eqnarray}
Bifurcations in the system (\ref{fast}) under variation of $\cc^{\rm s}$ as a parameter are
confronted to kinetic critical phenomena. Studies of more complicated dynamic phenomena in the fast
subsystem (\ref{fast}) require various techniques of averaging, stability analysis of the averaged
quantities etc.

Various versions of the QSS method are well possible, and are actually used widely, for example, the
hierarchical QSS method. There, one defines not a single fast subsystem but a hierarchy of them,
$\cc^{{\rm f}_1},\dots,\cc^{{\rm f}_k}$. Each subsystem  $\cc^{{\rm f}_i}$ is regarded as a slow
system for all the foregoing subsystems, and it is regarded as a fast subsystem for the following
members of the hierarchy. Instead of one system of equations (\ref{fast}), a hierarchy of systems of
lower-dimensional equations is considered, each of these subsystem is easier to study analytically.

Theory of singularly perturbed systems of ordinary differential equations is used to provide a
mathematical background and further development of the QSS approximation \cite{Bowen63,Segel89}. In
spite of a broad literature on this subject, it remains, in general, unclear, what is the smallness
parameter that separates the intermediate (fast) species from the basic (slow). Reaction rate
constants cannot be such a parameter (unlike in the case of the quasi-equilibrium). Indeed,
intermediate species participate in the {\it same} reactions, as the basic species (for example,
${\rm H}_2\rightleftharpoons 2{\rm H}$, ${\rm H}+{\rm O}_2\rightleftharpoons {\rm OH}+{\rm O}$). It
is therefore incorrect to state that $\cc^{\rm f}$ evolve faster than $\cc^{\rm s}$. In the sense of
reaction rate constants, $\cc^{\rm f}$ is not faster.

For catalytic reactions, it is not difficult to figure out what is the smallness parameter that
separates the intermediate species from the basic, and which allows to upgrade the QSS assumption to
a singular perturbation theory rigorously \cite{YBGE91}. This smallness parameter is the ratio of
balances: Intermediate species include the catalyst, and their total amount is simply significantly
less than the amount of all the $\cc_i$'s. After renormalizing to the variables of one order of
magnitude, the small parameter appears explicitly.

For usual radicals, the origin of the smallness parameter is quite similar. There are much less
radicals than the basic species (otherwise, the QSS assumption is inapplicable). In the case of
radicals, however, the smallness parameter cannot be extracted directly from balances $B_i$
(\ref{conser}). Instead, one can come up with a thermodynamic estimate: Function $G$ decreases in the
course of reactions, whereupon we obtain the limiting estimate of concentrations of any specie:
\begin{equation}
\label{TDlim} c_i\le \max_{G(\cc)\le G(\cc(0))} c_i,
\end{equation}
where $\cc(0)$ is the initial composition. If the concentration $c_{\rm R}$ of the radical R is small
both initially and in the equilibrium, then it should remain small also along the path to the
equilibrium. For example, in the case of ideal $G$ (\ref{gfun}) under relevant conditions, for any
$t>0$, the following inequality is valid:
\begin{equation}
\label{INEQ_R} c_{\rm R}[\ln(c_{\rm R}(t)/c_{\rm R}^{\rm eq})-1]\le G(\cc(0)).
\end{equation}
Inequality (\ref{INEQ_R}) provides the simplest (but rather coarse) thermodynamic estimate of $c_{\rm
R}(t)$ in terms of $G(\cc(0))$ and $c_{\rm R}^{\rm eq}$ {\it uniformly for $t>0$}. Complete theory of
thermodynamic estimates of dynamics has been developed in \cite{G84}. One can also do computations
without a priori estimations, if one accepts the QSS assumption until the values $\cc^{\rm f}$ stay
sufficiently small.

Let us assume that an a priori estimate has been found, $c_i(t)\le c_{i\ {\rm max}}$, for each $c_i$.
These estimate may depend on the initial conditions, thermodynamic data etc. With these estimates, we
are able to renormalize the variables in the kinetic equations (\ref{reaction}) in such a way that
renormalized variables take their values from the unit segment $[0,1]$: $\tilde{c}_i=c_i/c_{i\ {\rm
max}}$. Then the system (\ref{reaction}) can be written as follows:
\begin{equation}
\label{reduced} \frac{d\tilde{c}_i}{dt}=\frac{1}{c_{i\ {\rm max}}}J_i(\cc).
\end{equation}
The system of dimensionless parameters, $\epsilon_i=c_{i\ {\rm max}}/\max_i c_{i\ {\rm max}}$ defines
a hierarchy of relaxation times, and with its help one can establish various realizations of the QSS
approximation. The simplest version is the standard QSS assumption: Parameters $\epsilon_i$ are
separated in two groups, the smaller ones, and of the order $1$. Accordingly, the concentration
vector is split into $\cc^{\rm s}\oplus\cc^{\rm f}$. Various hierarchical QSS are possible, with
this, the problem becomes more tractable analytically.

Corrections to the QSS approximation can be addressed in various ways (see, e.\ g.,
\cite{Vasil'eva95,Strygin88}). There exist a variety of ways to introduce the smallness parameter
into kinetic equations, and one can find applications to each of the realizations. However, the two
particular realizations remain basic for chemical kinetics: (i) Fast reactions (under a given
thermodynamic data); (ii) Small concentrations. In the first case, one is led to the
quasi-equilibrium approximation, in the second case - to the classical QSS assumption. Both of these
approximations allow for hierarchical realizations, those which include not just two but many
relaxation time scales. Such a multi-scale approach {\it essentially simplifies} analytical studies
of the problem.

The method of invariant manifold which we present below in the section \ref{MIM} allows to use both
the QE and the QSS as initial approximations in the iterational process of seeking slow invariant
manifolds. It is also possible to use a  different initial ansatz chosen by a physical intuition,
like, for example, the Tamm--Mott-Smith approximation in the theory of strong shock waves
\cite{GK92}.

\subsection{Methods based on spectral decomposition of Jacobian fields}\label{Jacobi}

The idea to use the spectral decomposition of Jacobian fields in the problem of separating the
motions into fast and slow originates from methods of analysis of stiff systems \cite{Gear71}, and
from methods of sensitivity analysis in control theory \cite{Rabitz83}. There are two basic
statements of the problem for these methods: (i) The problem of the slow manifold, and (ii) The
problem of a complete decomposition (complete integrability) of kinetic equations. The first of these
problems consists in constructing the slow manifold $\bOmega$, and a decomposition of motions into
the fast one - towards $\bOmega$, and the slow one - along $\bOmega$ \cite{Maas92}. The second of
these problems consists in a transformation of kinetic equations (\ref{reaction}) to a diagonal form,
$\dot{\zeta}_i=f_i(\zeta_i)$  (so-called {\it full nonlinear lumping}
 or {\it modes decoupling}, \cite{Lam94,Li94,Toth97}).
 Clearly, if one finds a sufficiently explicit solution
to the second problem, then  the system (\ref{reaction}) is completely integrable, and nothing more
is needed, the result has to be simply used. The question is only to what extend such a solution can
be possible, and how difficult it would be as compared to the first problem to find it.

One of the currently  most popular methods is the construction of the so-called {\it intrinsic
low-dimensional manifold} (ILDM, \cite{Maas92}). This method is based on the following geometric
picture: For each point $\cc$, one defines the Jacobian matrix of Eq.\ (\ref{reaction}),
$\FF_{\cc}\equiv \partial \JJ(\cc)/\partial\cc$. One assumes that, in the domain of interest, the
eigenvalues of $\FF_{\cc}$ are separated into two groups, $\lambda_i^{\rm s}$ and $\lambda_j^{\rm
f}$, and that the following inequalities are valid:
\[ {\rm Re}\ \lambda_i^{\rm s}\ge a > b\ge {\rm Re} \lambda_j^{\rm f},\ a\gg b,\ b<0.\]
Let us denote as $L_{\cc}^{\rm s}$ and $L_{\cc}^{\rm f}$ the invariant subspaces corresponding to
$\lambda^{\rm s}$ and $\lambda^{\rm f}$, respectively, and let $\ZZ_{\cc}^{\rm s}$ and
$\ZZ_{\cc}^{\rm f}$ be the corresponding spectral projectors, $\ZZ_{\cc}^{\rm s}L_{\cc}^{\rm
s}=L_{\cc}^{\rm s}$, $\ZZ_{\cc}^{\rm f}L_{\cc}^{\rm f}=L_{\cc}^{\rm f}$, $\ZZ_{\cc}^{\rm
s}L_{\cc}^{\rm f}=\ZZ_{\cc}^{\rm f}L_{\cc}^{\rm s}=\{0\}$, $\ZZ_{\cc}^{\rm s}+\ZZ_{\cc}^{\rm f}=1$.
Operator $\ZZ_{\cc}^{\rm s}$ projects onto the subspace of ``slow modes'' $L_{\cc}^{\rm s}$, and it
annihilates the ``fast modes'' $L_{\cc}^{\rm f}$. Operator $\ZZ_{\cc}^{\rm f}$ does the opposite, it
projects onto fast modes, and it annihilates the slow modes. The basic equation of the ILDM reads:
\begin{equation}
\label{ILDM} \ZZ_{\cc}^{\rm f}\JJ(\cc)=0.
\end{equation}
In this equation, the unknown is the concentration vector $\cc$. The set of solutions to Eq.\
(\ref{ILDM}) is the ILDM manifold $\bOmega_{{\rm ildm}}$.

For linear systems, $\FF_{\cc}$, $\ZZ_{\cc}^{\rm s}$, and $\ZZ_{\cc}^{\rm f}$, do not depend on
$\cc$, and $\bOmega_{{\rm ildm}}=\cc^{\rm eq}+L^{\rm s}$. On the other hand, obviously, $\cc^{\rm
eq}\in\bOmega_{{\rm ildm}}$. Therefore, procedures of solving of Eq.\ (\ref{ILDM}) can be initiated
by choosing the linear approximation, $\bOmega_{{\rm ildm}}^{(0)}=\cc^{\rm eq}+L^{\rm s}_{\cc^{\rm
eq}}$, in the neighborhood of the equilibrium $\cc^{\rm eq}$, and then continued parametrically into
the nonlinear domain. Computational technologies of a continuation of solutions with respect to
parameters are well developed (see, for example, \cite{Khibnik93,Roose90}). The problem of the
relevant parameterization is solved locally: In the neighborhood of a given point $\cc^0$ one can
choose $\ZZ_{\cc}^{\rm s}(\cc-\cc^0)$ for a characterization of the vector $\cc$. In this case, the
space of parameters is $L_{\cc}^{\rm s}$. There exist other, physically motivated ways to
parameterize manifolds (\cite{GK92}; see also section \ref{TDP} below).

There are two drawbacks of the ILDM method which call for its refinement: (i) {\it ``Intrinsic'' does
not imply ``invariant''.} Eq.\ (\ref{ILDM}) is not invariant of the dynamics (\ref{reaction}). If one
differentiates Eq.\ (\ref{ILDM}) in time due to Eq.\ (\ref{reaction}), one obtains a new equation
which is the implication of Eq.\ (\ref{ILDM}) {\it only} for linear systems. In a general case, the
motion $\cc(t)$ takes off the $\bOmega_{{\rm ildm}}$. Invariance of a manifold $\bOmega$ means that
$\JJ(\cc)$ touches $\bOmega$ in every point $\cc\in\bOmega$. It remains unclear how the ILDM
(\ref{ILDM}) corresponds with this condition. Thus, from the dynamical perspective, the status of the
ILDM remains not well defined, or ``ILDM is ILDM'', defined self-consistently by Eq.\ (\ref{ILDM}),
and that is all what can be said about it. (ii) From the geometrical standpoint, spectral
decomposition of Jacobian fields is not the most attractive way to compute manifolds. If we are
interested in the behavior of trajectories, how they converge or diverge, then one should consider
the symmetrized part of $\FF_{\cc}$, rather than $\FF_{\cc}$ itself.

Symmetric part, $\FF_{\cc}^{\rm sym}=(1/2)(\FF_{\cc}^{\dag}+\FF_{\cc})$, defines the dynamics of the
distance between two solutions,  $\cc$ and $\cc'$,
 in a given local Euclidean metrics.
Skew-symmetric part defines rotations. If we want to study manifolds based on the argument about
convergence/divergence of trajectories, then we should use in Eq.\ (\ref{ILDM}) the spectral
projector $\ZZ_{\cc}^{\rm f sym}$ for the operator $\FF_{\cc}^{\rm sym}$. This, by the way, is  also
a  significant simplification from the standpoint of computations. It remains to choose the metrics.
This choice is unambiguous from the thermodynamic perspective. In fact, there is only one choice
which fits into the physical meaning of the problem, this is the metrics associated with the
thermodynamic (or entropic) scalar product,
\begin{equation}
\label{ESP} \langle\xxx,\yy\rangle=(\xxx,\HH_{\cc}\yy),
\end{equation}
where $\HH_{\cc}$ is the matrix of second-order derivatives of $G$ (\ref{MATRIX}). In the
equilibrium, operator $\FF_{\cc^{\rm eq}}$ is selfadjoint with respect to this scalar product
(Onsager's reciprocity relations). Therefore, the behavior of the ILDM in the vicinity of the
equilibrium does not alter under the replacement, $\FF_{\cc^{\rm eq}}=\FF_{\cc^{\rm eq}}^{\rm sym}$.
In terms of usual matrix representation, we have:
\begin{equation}
\FF_{\cc}^{\rm sym}=\frac{1}{2}(\FF_{\cc}+\HH_{\cc}^{-1}\FF_{\cc}^{T}\HH_{\cc}),\label{Fsym}
\end{equation}
where $\FF_{\cc}^{T}$ is the ordinary transposition.

The ILDM constructed with the help of the symmetrized Jacobian field will be termed the {\it
symmetric entropic intrinsic low-dimensional manifold} (SEILDM). Selfadjointness of $\FF_{\cc}^{\rm
sym}$ (\ref{Fsym}) with respect to the thermodynamic scalar product (\ref{ESP}) simplifies
considerably computations of
 spectral decomposition. Moreover, it becomes necessary to do spectral decomposition
in only one point - in the equilibrium. Perturbation theory for selfadjoint operators is a very well
developed subject \cite{Kato76}, which makes it possible to easily extend the spectral decomposition
with respect to parameters. A more detailed discussion of the selfadjoint linearization will be given
below in section \ref{SA}.

Thus, when the geometric picture behind the decomposition of motions is specified, the physical
significance of the ILDM becomes more transparent, and it leads to its modification into the SEILDM.
This also gains simplicity in the implementation by switching from non-selfadjoint spectral problems
to selfadjoint. The quantitative estimate of this simplification is readily available: Let $d$ be the
dimension of the phase space, and $k$ the dimension of the ILDM ($k={\rm dim} L_{\cc}^{\rm s}$). The
space of all the projectors $\ZZ$ with the $k$-dimensional image has the dimension $D=2k(d-k)$. The
space of all the selfadjoint projectors with the $k$-dimensional image has the dimension $D^{\rm
sym}=k(d-k)$. For $d=20$ and $k=3$, we have $D=102$ and $D^{\rm sym}=51$. When the spectral
decomposition by means of parametric extension is addressed, one considers equations of the form:
\begin{equation}
\label{parametric} \frac{d\ZZ_{\cc(\tau)}^{\rm s}}{d\tau}=\bPsi^{\rm s}\left(\frac{d\cc}{d\tau},
\ZZ_{\cc(\tau)}^{\rm s}, \FF_{\cc(\tau)}, \bnabla\FF_{\cc(\tau)}\right),
\end{equation}
where $\tau$ is the parameter, and $\bnabla\FF_{\cc}=\bnabla\bnabla\JJ(\cc)$ is the differential of
the Jacobian field. For the selfadjoint case, where we use $=\FF_{\cc}^{\rm sym}$ instead of
 $\FF_{\cc}$, this system of equations
has twice less independent variables, and also the right hand is of a simpler structure.

It is more difficult to improve on  the first of the remarks (ILDM is not invariant). The following
naive approach may seem possible:

(i) Take $\bOmega_{\rm ildm}=\cc^{\rm eq}+L^{\rm s}_{\cc^{\rm eq}}$ in a neighborhood $U$ of the
equilibrium $\cc^{\rm eq}$. [This is also a useful initial approximation for solving Eq.\
(\ref{ILDM})].

(ii) Instead of computing the solution to Eq.\ (\ref{ILDM}), integrate the kinetic equations
(\ref{reaction}) {\it backwards in the time}. It is sufficient to take initial conditions $\cc(0)$
from a dense set on the boundary, $\partial U\cap (\cc^{\rm eq}+L^{\rm s}_{\cc^{\rm eq}})$, and to
compute solutions $\cc(t)$, $t<0$.

(iii) Consider the obtained set of trajectories as an approximation of the slow invariant manifold.

This approach will guarantee invariance, by construction, but it is prone to pitfalls in what
concerns the slowness. Indeed, the integration backwards in the time will see exponentially divergent
trajectories, if they were exponentially converging in the normal time progress. This way one finds
{\it some} invariant manifold which touches $\cc^{\rm eq}+L^{\rm s}_{\cc^{\rm eq}}$ in the
equilibrium. Unfortunately, there are infinitely many such manifolds, and they fill out almost all
the space of concentrations. However, we must select the slow component of motions. Such a
regularization is possible. Indeed, let us replace in Eq.\ (\ref{reaction}) the vector field
$\JJ(\cc)$ by the vector field $\ZZ_{\cc}^{{\rm s sym}}\JJ(\cc)$, and obtain a regularized kinetic
equation,
\begin{equation}
\label{regreaction} \dot{c}=\ZZ_{\cc}^{{\rm s sym}}\JJ(\cc).
\end{equation}
Let us replace integration backwards in time of the kinetic equation (\ref{reaction}) in the naive
approach described above by integration backwards in time of the regularized kinetic equation
(\ref{regreaction}). With this, we obtain a rather convincing version of the ILDM (SEILDM). Using
Eq.\ (\ref{parametric}), one also can write down an equation for the projector $\ZZ_{\cc}^{{\rm s
sym}}$, putting $\tau=t$. Replacement of Eq.\ (\ref{reaction}) by Eq.\ (\ref{regreaction}) also makes
the integration backwards in time in the naive approach more stable. However, {\it regularization
will again conflict with invariance}. The ``naive refinement'' after the regularization
(\ref{regreaction}) produces just a slightly  different  version of the ILDM (or SEILDM) but it does
not construct the slow invariant manifold. So, where is the way out? We believe that the ILDM and its
version SEILDM are, in general, good initial approximations of the slow manifold. However, if one is
indeed interested in finding the invariant manifold, one has to write out the true condition of
invariance and solve it. As for the initial approximation for the method of invariant manifold one
can use any ansatz, in particular, the SEILDM.

{\it The problem of a complete decomposition} of kinetic equations can be solved indeed in some
cases. The first such solution was the spectral decomposition for linear systems \cite{Wei62}.
Decomposition is sometimes possible also for nonlinear systems (\cite{Li94}; \cite{Toth97}). The most
famous example of a complete decomposition of infinite-dimensional kinetic equation is the complete
integrability of the space-independent Boltzmann equation for Maxwell`s molecules found in
\cite{Bobylev88}. However, in a general case, there exist no analytical, not even a twice
differentiable transformation which would decouple modes. The well known Grobman-Hartman theorem
\cite{Hartman63,Hartman82} states only the existence of a continuous transform which decomposes modes
in a neighborhood of the equilibrium. For example, the analytic planar system, $dx/dt=-x$,
$dy/dt=-2y+x^2$, is not $\CC^2$ linearizable. These problems remain of interest
 \cite{Chicone00}.
Therefore, in particular, it becomes quite ineffective to construct such a transformation in a form
of a series. It is more effective to solve a simpler problem of extraction of a slow invariant
manifold \cite{Beyn98}.

{\it Sensitivity analysis} \cite{Rabitz83,Rabitz87,Lam94} makes it possible to select essential
variables and reactions, and to decompose motions into fast  and slow. In a sense, the ILDM method is
a development of the sensitivity analysis. In particular, the {\it computational singular
perturbation} (CSP) method of \cite{Lam94} includes ILDM (or any other reasonable initial choice of
the manifold) into a procedure of consequent refinements. Recently, a further step in this direction
was done in \cite{Zhu99}. In this work, the authors use a {\it nonlocal in time criterion of
closeness of solutions} of the full and of the reduced systems of chemical kinetics. They require not
just a closeness of derivatives but a true closeness of the dynamics.

Let us be interested in the dynamics of the concentrations of just a few species, ${\rm A}_1,\dots,
{\rm A}_{p}$, whereas the rest of the species, ${\rm A}_{p+1}, \dots, {\rm A}_{n}$ are used for
building the kinetic equation, and for understanding the process. Let $\cc_{\rm goal}$ be the
concentration vector with components $c_1,\dots, c_p$, $\cc_{\rm goal}(t)$ be the corresponding
components of the solution to Eq.\ (\ref{reaction}), and $\cc^{\rm red}_{\rm goal}$ be the solution
to the simplified model with corresponding initial conditions. \cite{Zhu99} suggest to minimize the
difference between
 $\cc_{\rm goal}(t)$ and  $\cc^{\rm red}_{\rm goal}$ on the
segment $t\in[0,T]$: $\|\cc_{\rm goal}(t)-\cc^{\rm red}_{\rm goal}\|\to\min$. In the course of the
optimization under certain restrictions one selects the optimal (or appropriate) reduced model. The
sequential quadratic programming method and heuristic rules of sorting the reactions, substances etc
were used. In the result, for some stiff systems studied, one avoids typical pitfalls of the local
sensitivity analysis. In simpler situations this method should give similar results as the local
methods.

\subsection{Thermodynamic criteria for selection of important reactions}

One of the problems addressed by the sensitivity analysis is the selection of the important and
discarding the unimportant reactions. \cite{BYA77} suggested a simple principle to compare importance
of different reactions according to their contribution to the entropy production (or, which is the
same, according to their contribution to $\dot{G}$). Based on this principle, \cite{Dimitrov82}
described domains of parameters in which the reaction of hydrogen oxidation, ${\rm H}_2+{\rm
O}_2+{\rm M}$, proceeds due to different mechanisms. For each elementary reaction, he has derived the
domain inside which the contribution of this reaction is essential (nonnegligible). Due to its
simplicity, this entropy production principle is especially well suited for analysis of complex
problems. In particular, recently, a version of the entropy production principle was used in the
problem of selection of boundary conditions for Grad's moment equations \cite{Struchtrup98,GKZ02}.
For ideal systems (\ref{gfun}), the contribution of the $s$th reaction to $\dot{G}$ has a
particularly simple form:
\begin{equation}
\label{dotGs} \dot{G}_{s}=-W_s\ln\left(\frac{W_s^+}{W_s^-}\right),\ \dot{G}=\sum_{s=1}^{r}\dot{G}_s.
\end{equation}
For nonideal systems, the corresponding expressions (\ref{HMDD})
 are also not too complicated.

\section{Outline of the method of invariant manifold}
\label{MIMG}

In many cases, dynamics of the $d$-dimensional system (\ref{reaction}) leads to a manifold of a lower
dimension. Intuitively, a typical phase trajectory behaves as follows: Given the initial state
$\cc(0)$ at $t=0$, and after some period of time, the trajectory comes close to some low-dimensional
manifold $\bOmega$, and after that proceeds towards the equilibrium essentially along this manifold.
The goal is to construct this manifold.

The starting point of our approach is based on a formulation of the two main requirements:

(i). {\it Dynamic invariance}:  The manifold $\bOmega$ should  be (positively)  invariant  under the
dynamics  of the originating system (\ref{reaction}): If $\cc(0)\in\bOmega$, then $\cc(t)\in\bOmega$
for each $t>0$.

(ii). {\it Thermodynamic consistency of the reduced dynamics}: Let  {\it some} (not obligatory
invariant) manifold $\bOmega$ is considered as a manifold of reduced description. We should define a
set of linear operators, $\PP_{\cc}$, labeled by the states $\cc\in\bOmega$, which project the
vectors $\JJ(\cc)$, $\cc\in\bOmega$ onto the tangent bundle of the manifold $\bOmega$, thereby
generating the induced vector field, $\PP_{\cc}\JJ(\cc)$, $\cc\in\bOmega$. This induced vector field
on the tangent bundle of the manifold $\bOmega$ is identified with the reduced dynamics along the
manifold $\bOmega$. The thermodynamicity requirement for this induced vector field reads
\begin{equation}
\label{thermo} (\bnabla G(\cc),\PP_{\cc}\JJ(\cc))\leq 0,\mbox{ for\ each\ }\cc\in\bOmega.
\end{equation}

In order to meet these requirements, the method of invariant manifold suggests two complementary
procedures:

(i). To treat the condition of dynamic invariance as an equation, and to solve it iteratively by a
Newton method. This procedure is geometric in its nature, and it does not use the time dependence and
small parameters.

(ii). Given an approximate manifold of reduced description, to construct the projector satisfying the
condition (\ref{thermo}) in a way which does not depend on the vector field $\JJ$.

We shall now outline both these procedures starting with the second. The solution consists, in the
first place, in formulating the {\it thermodynamic condition} which should be met by the projectors
$\PP_{\cc}$: For each $\cc\in\bOmega$, let us consider the linear functional
\begin{equation}
\label{FUNC} M^*_{\cc}(\xx)=(\bnabla G(\cc),\xx).
\end{equation}
Then the thermodynamic condition for the projectors reads:
\begin{equation}
\label{therm1} {\rm ker}\PP_{\cc}\subseteq{\rm ker}M^*_{\cc},\ {\rm  for\ each}\ \cc\in\bOmega.
\end{equation}
Here ${\rm ker}\PP_{\cc}$ is the null space of the projector, and ${\rm ker}M^*_{\cc}$ is the
hyperplane orthogonal to the vector $M^*_{\cc}$. It has been shown  \cite{GK92,GK94} that the
condition (\ref{therm1}) is the necessary and sufficient condition to establish the thermodynamic
induce vector field on the given manifold $\bOmega$ for all possible dissipative vector fields $\JJ$
simultaneously.

Let us now turn to the requirement of invariance. By a definition, the manifold  $\bOmega$  is
invariant with respect to the vector field $\JJ$  if and only if the following equality is true:
\begin{equation}
\label{INVARIANCE} \left[1-\PP\right]\JJ(\cc)=0,\ {\rm for\ each\ } \cc\in\bOmega.
\end{equation}
In this expression $\PP$ is an {\it arbitrary} projector on the tangent bundle of the manifold
$\bOmega$. It has been suggested to consider the condition (\ref{INVARIANCE}) as an {\it equation} to
be solved iteratively starting with some appropriate initial manifold.

Iterations for the invariance equation (\ref{INVARIANCE}) are considered  in the section \ref{DC}.
The next section presents construction of the thermodynamic projector using a specific
parameterization of manifolds.

\section{Thermodynamic projector}
\label{THERMO}

\subsection{Thermodynamic parameterization}
\label{TDP}

In this section, $\bOmega$ denotes a generic $p$--dimensional manifold. First, it should be mentioned
that {\it any} parameterization of $\bOmega$ generates a certain projector, and thereby a certain
reduced dynamics. Indeed, let us consider a set of $m$ independent functionals
$M(\cc)=\{M_1(\cc),\dots,M_p(\cc)\}$, and let us assume that they form a coordinate system on
$\bOmega$ in such a way that $\bOmega=\cc(M)$, where $\cc(M)$ is a vector function of the parameters
$M_1,\dots,M_p$. Then the projector associated with this parameterization reads:
\begin{equation}
\label{proj} \PP_{\cc(M)}\xx=\sum_{i=1}^p\frac{\partial\cc(M)}{\partial M_i} (\bnabla
M_i\bigm|_{\cc(M)},\xx),
\end{equation}
where $N^{-1}_{ij}$ is the inverse to the $p\times p$ matrix:
\begin{equation}
\label{NORMALIZATION} \NN(M)=\|( \bnabla M_i,\partial\cc/\partial M_j)\|.
\end{equation}
This somewhat involved notation is intended to stress that the projector (\ref{proj}) is dictated by
the choice of the parameterization. Subsequently, the induced vector field of the reduced dynamics is
found by applying projectors (\ref{proj}) on the vectors $\JJ(\cc(M))$, thereby inducing the reduced
dynamics in terms of the parameters $M$ as follows:
\begin{equation}
\label{dyn} \dot{M}_i=\sum_{j=1}^pN^{-1}_{ij}(M) (\bnabla M_j\bigm|_{\cc(M)},\JJ(\cc(M))),
\end{equation}
Depending on the choice of the parameterization, dynamic equations (\ref{dyn}) are (or are not)
consistent with the thermodynamic requirement (\ref{thermo}). The {\it thermodynamic
parameterization} makes use of the condition (\ref{therm1}) in order to establish the thermodynamic
projector. Specializing to the case (\ref{proj}), let us consider the linear functionals,
\begin{equation}
\label{derivative} DM_i\bigm|_{\cc(M)}(\xx)= (\bnabla M_i\bigm|_{\cc(M)},\xx).
\end{equation}
Then the condition (\ref{therm1}) takes the form:
\begin{equation}
\label{therm2} \bigcap_{i=1}^p{\rm ker}DM_i\bigm|_{\cc(M)}\subseteq{\rm ker} M^*_{\cc(M)},
\end{equation}
that is, the intersection of null spaces of the functionals (\ref{derivative}) should belong to the
null space of the differential of the Lyapunov function $G$, in each point of the manifold $\bOmega$.

In practice, in order to construct the thermodynamic parameterization, we take the following set of
functionals in each point $\cc$ of the manifold $\bOmega$:
\begin{eqnarray}
\label{param_thermo} M_1(\xx)&=&M_{\cc}^*(\xx),\ \cc\in\bOmega\\ M_i(\xx)&=&(\mm_i,\xx),\ i=2,\dots,p
\label{param_other}
\end{eqnarray}
It is required that vectors $\bnabla G(\cc), \mm_2,\dots,\mm_p$ are linearly independent in each
state $\cc\in\bOmega$. Inclusion of the functionals (\ref{FUNC}) as a part of the system
(\ref{param_thermo}) and (\ref{param_other}) implies the thermodynamic condition (\ref{therm2}).
Also, any linear combination of the parameter set (\ref{param_thermo}), (\ref{param_other})  will
meet the thermodynamicity requirement.

It is important to notice here that the thermodynamic condition is satisfied whatsoever the
functionals $M_2,\dots,M_p$ are. This is very convenient for it gives an opportunity to take into
account the conserved quantities correctly. The manifolds we are going to deal with should be
consistent with the conservation laws (\ref{conser}). While the explicit characterization of the
phase space $\VV$ is a problem on its own, in practice, it is customary to work in the
$n$--dimensional space while keeping the constraints (\ref{conser}) explicitly on each step of the
construction. For this technical reason, it is convenient to consider manifolds of the dimension
$p>l$, where $l$ is the number of conservation laws, in the $n$--dimensional space rather than in the
phase space $\VV$.  The thermodynamic parameterization is then concordant also with the conservation
laws if $l$ of the linear functionals (\ref{param_other}) are identified with the conservation laws.
In the sequel, only projectors consistent with conservation laws are considered.

Very frequently, the manifold $\bOmega$ is represented as a $p$-parametric family
$\cc(a_1,\dots,a_p)$, where $a_i$ are coordinates on the manifold. The thermodynamic {\it
re-parameterization} suggests a representation of the coordinates $a_i$ in terms of $M_{\cc}^*,
M_2,\dots,M_p$ (\ref{param_thermo}), (\ref{param_other}). While the explicit construction of these
functions may be a formidable task, we notice that the construction of the thermodynamic projector of
the form (\ref{proj}) and of the dynamic equations (\ref{dyn}) is relatively easy because only the
derivatives $\partial\cc/\partial M_i$ enter these expressions. This point was discussed in a detail
in  \cite{GK92,GK94}.

\subsection{Decomposition of motions: Thermodynamics}
\label{DEC_THERM}

Finally, let us discuss how the thermodynamic projector is related to the decomposition of motions.
{\it Assuming} that the decomposition of motions near the manifold $\bOmega$ is true indeed, let us
consider states which were initially close enough to the manifold $\bOmega$. Even without knowing the
details about the evolution of the states towards $\bOmega$, we know that the Lyapunov function $G$
was decreasing in the course of this evolution. Let us consider a set of states $\UU_{\cc}$
 which contains all those vectors $\cc'$ that
have arrived (in other words, have been projected) into the point $\cc\in\bOmega$. Then we observe
that the state $\cc$ furnishes the minimum of  the function $G$ on the set $\UU_{\cc}$.
 If a state $\cc'\in\UU_{\cc}$, and if it deviates small enough
from the state $\cc$ so that the linear approximation is valid, then $\cc'$ belongs to the affine
hyperplane
\begin{equation}
\label{hyperplane} \Gamma_{\cc}=\cc+{\rm ker\ }M_{\cc}^*,\ \cc\in\bOmega.
\end{equation}
This hyperplane actually participates in the condition (\ref{therm1}). The consideration was entitled
`thermodynamic' \cite{GK92} because it describes the states $\cc\in\bOmega$ as points of minimum of
the function $G$ over the corresponding hyperplanes (\ref{hyperplane}).

\section{Corrections}
\label{DC}

\subsection{Preliminary discussion}

The thermodynamic projector is needed to induce the dynamics on a given manifold in such a way that
the dissipation inequality (\ref{thermo}) holds. Coming back to the issue of constructing
corrections, we should stress that the projector participating in the invariance condition
(\ref{INVARIANCE}) is arbitrary. It is convenient to make use of this point: When Eq.\
(\ref{INVARIANCE}) is solved iteratively, the projector may be kept non--thermodynamic unless the
induced dynamics is explicitly needed.

Let us assume that we have chosen the initial manifold, $\bOmega_0$, together with the associated
projector $\PP_0$, as the first approximation to the desired manifold of reduced description. Though
the choice of the initial approximation $\bOmega_0$ depends on the specific problem, it is often
reasonable to consider quasi-equilibrium or quasi steady-state approximations. In most cases, the
manifold $\bOmega_0$ is not an invariant manifold. This means that $\bOmega_0$ does not satisfy the
invariance condition (\ref{INVARIANCE}):
\begin{equation}
\label{DEFECT} \bDelta_0=[1-\PP_0]\JJ(\cc_0)\ne0,\ {\rm for\ some\ } \cc_0\in\bOmega_0.
\end{equation}
Therefore, we seek a correction $\cc_1=\cc_0+\delta\cc$. Substituting $\PP=\PP_0$ and
$\cc=\cc_0+\delta\cc$ into the invariance equation (\ref{INVARIANCE}), and after the linearization in
$\delta\cc$, we derive the following linear equation:
\begin{equation}
\label{METHOD} \left[1-\PP_{0}\right]\left[\JJ(\cc_0)+ \LL_{\cc_0}\delta\cc\right]=0,
\end{equation}
where $\LL_{c_0}$ is the matrix of first derivatives of the vector function $\JJ$, computed in the
state $\cc_0\in\bOmega_0$. The system of linear algebraic equations (\ref{METHOD}) should be supplied
with the additional condition.
\begin{equation}
\label{uni_con} \PP_0\delta\cc=0.
\end{equation}

In order to illustrate the nature of the Eq.\ (\ref{METHOD}), let us consider the case of linear
manifolds for linear systems. Let a linear evolution equation is given in the finite-dimensional real
space: $\dot{\cc}=\LL\cc$, where $\LL$ is negatively definite symmetric matrix with a simple
spectrum. Let us further assume the quadratic Lyapunov function, $G(\cc)=(\cc,\cc)$. The manifolds we
consider are lines, $\vl(a)=a\ee$, where $\ee$ is the unit vector, and $a$ is a scalar. The
invariance equation for such manifolds reads: $\ee(\ee,\LL\ee)-\LL\ee=0$, and is simply a form of the
eigenvalue problem for the operator $\LL$. Solutions to the latter equation are eigenvectors
$\ee_{i}$, corresponding to eigenvalues $\lambda_{i}$.

Assume that we have chosen a line, $\vl_0=a\ee_0$, defined by the unit vector $\ee_0$, and that $e_0$
is not an eigenvector of $\LL$. We seek another line, $\vl_1=a\ee_1$, where $\ee_1$ is another unit
vector, $\ee_1=\yy_1/\|\yy_1\|$, $\yy_1=\ee_0+\delta\yy$. The additional condition (\ref{uni_con})
now reads: $(\delta\yy,\ee_0)=0$. Then the Eq.\ (\ref{METHOD}) becomes
$[1-\ee_0(\ee_0,\cdot)]L[\ee_0+\delta\yy]=0$. Subject to the additional condition, the unique
solution is as follows: $\ee_0+\delta\yy= (\ee_0,\LL^{-1}\ee_0)^{-1}\LL^{-1}\ee_0$. Rewriting the
latter expression in the eigen--basis of $\LL$, we have: $\ee_0+\delta\yy\propto
\sum_{i}\lambda_i^{-1}\ee_i(\ee_i,\ee_0)$. The leading term in this sum corresponds to the eigenvalue
with the minimal absolute value. The example indicates that the  method of linearization
(\ref{METHOD}) seeks the direction of the {\it slowest relaxation}. For this reason, the method
(\ref{METHOD}) can be recognized as the basis of an iterative method for constructing the manifolds
of slow motions.

For the nonlinear systems, the matrix $\LL_{c_0}$ in the Eq.\ (\ref{METHOD}) depends nontrivially on
$\cc_0$. In this case the system (\ref{METHOD}) requires a further specification which will be done
now.

\subsection{Symmetric linearization}
\label{SA}

The invariance condition (\ref{INVARIANCE}) supports a lot of invariant manifolds, and not all of
them are relevant to the reduced description (for example, any individual trajectory is itself an
invariant manifold). This should be carefully taken into account when deriving a relevant equation
for the correction in the states of the initial manifold $\bOmega_0$ which are located far from
equilibrium. This point concerns the procedure of the linearization of the vector field $\JJ$,
appearing in the equation (\ref{METHOD}). We shall return to the explicit form of the Marcelin--De
Donder kinetic function (\ref{MDD}). Let $\cc$ is an arbitrary fixed element of the phase space. The
linearization of the vector function $\JJ$ (\ref{KINETIC MDD}) about $\cc$ may be written
$\JJ(\cc+\delta\cc)\approx\JJ(\cc)+\LL_{\cc}\delta\cc$ where the linear operator $\LL_{\cc}$ acts as
follows:
\begin{equation}
\label{LINEAR} \LL_{\cc}\xx=\sum_{s=1}^{r} \bgamma_s[W^{+}_s(\cc) (\balpha_s,\HH_{\cc}\xx )-
W^{-}_s(\cc)(\bbeta_s,\HH_{\cc} \xx)].
\end{equation}
Here $\HH_{\cc}$ is the matrix of second derivatives of the function $G$ in the state $\cc$ [see Eq.\
(\ref{MATRIX})]. The matrix $\LL_{\cc}$ in the Eq.\ (\ref{LINEAR}) can be decomposed as follows:
\begin{equation}
\label{DECOMPOSITION} \LL_{\cc}=\LL'_{\cc}+ \LL''_{\cc}.
\end{equation}
Matrices $\LL'_{\cc}$ and $\LL''_{\cc}$ act as follows:
\begin{eqnarray}
\label{SYMMETRIC} \LL'_{\cc}\xx&=&-\frac{1}{2}\sum_{s=1}^{r} [W^{+}_s(\cc)+W^{-}_s(\cc)]
\bgamma_s(\bgamma_s, \HH_{\cc}\xx),\\ \label{NONSYMMETRIC} \LL''_{\cc}\xx&=&\frac{1}{2}\sum_{s=1}^{r}
[W^{+}_s(\cc)-W^{-}_s(\cc)]\bgamma_s (\balpha_s+\bbeta_s, \HH_{\cc}\xx).
\end{eqnarray}
Some features of this decomposition are best seen when we use the thermodynamic scalar product
(\ref{ESP}): The following properties of the matrix $\LL'_{\cc}$ are verified immediately:

(i) The matrix $\LL'_{\cc}$ is symmetric in the scalar product (\ref{ESP}):
\begin{equation}
\label{sym} \langle \xx,\LL'_{\cc}\yy \rangle = \langle\yy,\LL'_{\cc}\xx \rangle.
\end{equation}
(ii) The matrix $\LL'_{\cc}$ is nonpositive definite in the scalar product (\ref{ESP}):
\begin{equation}
\label{pos} \langle \xx,\LL'_{\cc}\xx \rangle \le 0.
\end{equation}
(iii) The null space of the matrix $\LL'_{\cc}$ is the linear envelope of the vectors
$\HH^{-1}_{\cc}\bb_i$ representing the complete system of conservation laws:
\begin{equation}
\label{bal} {\rm ker}\LL'_{\cc}={\rm Lin}\{\HH^{-1}_{\cc}\bb_i, i=1,\dots,l\}
\end{equation}
(iv)  If $\cc=\cc^{\rm eq}$, then $W_s^+(\cc^{\rm eq})=W_s^-(\cc^{\rm eq})$, and
\begin{equation}
\label{prop4} \LL'_{\cc^{\rm eq}}=\LL_{\cc^{\rm eq}}.
\end{equation}

Thus, the decomposition Eq.\ (\ref{DECOMPOSITION}) splits the matrix $\LL_{\cc}$ in two parts: one
part, Eq.\ (\ref{SYMMETRIC}) is symmetric and nonpositive definite, while the other part, Eq.\
(\ref{NONSYMMETRIC}), vanishes in the equilibrium. The decomposition Eq.\ (\ref{DECOMPOSITION})
explicitly takes into account the Marcelin-De Donder form of the kinetic function. For other
dissipative systems, the decomposition (\ref{DECOMPOSITION}) is possible as soon as the relevant
kinetic operator is written in a gain--loss form [for instance, this is straightforward for the
Boltzmann collision operator].

In the sequel, we shall make use of the properties of the operator $\LL'_{\cc}$ (\ref{SYMMETRIC}) for
constructing the dynamic correction by extending the picture of the decomposition of motions.

\subsection{Decomposition of motions: Kinetics}
\label{DEC_KIN}

The assumption about the existence of the decomposition of motions near the manifold of reduced
description $\bOmega$ has led to the {\it thermodynamic} specifications of the states
$\cc\in\bOmega$. This was accomplished in the section \ref{DEC_THERM}, where the thermodynamic
projector was backed by an appropriate variational formulation, and this helped us to establish the
induced dynamics consistent with the dissipation property. Another important feature of the
decomposition of motions is that the states $\cc\in\bOmega$ can be specified {\it kinetically}.
Indeed, let us do it again as if the decomposition of motions were valid in the neighborhood of the
manifold $\bOmega$, and let us
 `freeze' the slow dynamics along the $\bOmega$, focusing on the
fast process of relaxation towards a state $\cc\in\bOmega$. From the thermodynamic perspective, fast
motions take place on the affine hyperplane $\cc+\delta\cc\in\Gamma_{\cc_0}$, where $\Gamma_{\cc_0}$
is given by Eq.\ (\ref{hyperplane}). From the kinetic perspective, fast motions on this hyperplane
should be treated as a  {\it relaxation} equation, equipped with the quadratic Lyapunov function
$\delta G=\langle\delta\cc,\delta\cc\rangle$, Furthermore, we require that the linear operator of
this evolution equation
 should respect Onsager's symmetry
requirements (selfadjointness with respect to the entropic scalar product). This latter crucial
requirement describes fast motions under the frozen slow evolution in the similar way, as {\it all}
the  motions near the equilibrium.

Let us consider now the manifold $\bOmega_0$ which is not the invariant manifold of the reduced
description but, by our assumption, is located close to it. Consider a state $\cc_0\in\bOmega_0$, and
the states $\cc_0+\delta\cc$ close to it. Further, let us consider an equation
\begin{equation}
\label{RELAX_SYMM} \dot{\delta\cc}=\LL'_{\cc_0}\delta\cc.
\end{equation}
Due to the properties of the operator $\LL'_{\cc_0}$ (\ref{SYMMETRIC}), this equation can be regarded
as a model of the assumed true relaxation equation near the true manifold of the reduced description.
For this reason, we shall use the symmetric operator $\LL'_{\cc}$ (\ref{SYMMETRIC}) {\it instead} of
the linear operator $\LL_{\cc}$
 when constructing the corrections.

\subsection{Symmetric iteration}
Let the manifold $\bOmega_0$ and the corresponding projector $\PP_0$ are the initial approximation to
the invariant manifold of the reduced description. The dynamic correction $\cc_1=\cc_0+\delta\cc$ is
found upon solving the following system of linear algebraic equations:
\begin{equation}
\label{ITERATION} \left[1-\PP_0\right]\left[\JJ(\cc_0) + \LL'_{\cc_0}\delta\cc\right]=0,\
\PP_0\delta\cc=0.
\end{equation}
Here $\LL'_{\cc_0}$ is the matrix (\ref{SYMMETRIC}) taken in the states on the manifold $\bOmega_0$.
An important technical point here is that the linear system (\ref{ITERATION}) always has the unique
solution for any choice of the manifold $\bOmega$. This point is crucial since it guarantees the
opportunity of carrying out the correction process for arbitrary number of steps.

\section{The method of invariant manifold}
\label{MIM}

We shall now combine together the two procedures discussed above. The resulting method of invariant
manifold intends to seek iteratively the reduced description, starting with an initial approximation.

(i). {\it Initialization}. In order to start the procedure, it is required to choose the initial
manifold $\bOmega_0$, and to derive corresponding thermodynamic projector $\PP_0$. In the majority of
cases, initial manifolds are available in two different ways. The first case are the
quasi-equilibrium manifolds described in the section \ref{partial_eq}. The macroscopic parameters are
$M_i=c_i=(\mm_i,\cc)$, where $\mm_i$ is the unit vector corresponding to the specie $A_i$. The
quasi-equilibrium manifold,  $\cc_0(M_1,\dots,M_k,B_1,\dots,B_l)$, compatible with the conservation
laws, is the solution to the variational problem:
\begin{eqnarray}
\label{QE} G\to{\rm min\ }, && (\mm_i,\cc)=c_i, \ i=1,\dots,k,\\\nonumber &&(\bb_j,\cc)=B_j, \
j=1,\dots,l .
\end{eqnarray}
In the case of quasi--equilibrium approximation, the corresponding thermodynamic projector can be
written most straightforwardly in terms of the variables $M_i$:
\begin{equation}
\label{PROJ-QE} \PP_0\xx=\sum_{i=1}^{k}\frac{\partial\cc_0}{\partial c_i} (\mm_i,\xx)+
\sum_{i=1}^{l}\frac{\partial\cc_0}{\partial B_i} (\bb_i,\xx).
\end{equation}
For quasi-equilibrium manifolds, a reparameterization with the set (\ref{param_thermo}),
(\ref{param_other}) is {\it not} necessary (\cite{GK92}; \cite{GK94}).

The second source of initial approximations are  quasi-stationary manifolds (section \ref{QSS}).
Unlike the quasi-equilibrium case, the quasi-stationary manifolds must be reparameterized in order to
construct the thermodynamic projector.

(ii). {\it Corrections.} Iterations are organized in accord with the rule: If $\cc_m$ is the $m$th
approximation to the invariant manifold, then the correction $\cc_{m+1}=\cc_{m}+\delta\cc$ is found
from the linear algebraic equations,
\begin{eqnarray}
\label{corr_eq} [1-\PP_{m}](\JJ(\cc_m)+\LL'_{\cc_m}\delta\cc)&=&0,\\
\PP_m\delta\cc&=&0.\label{corr_cond}
\end{eqnarray}
Here $\LL'_{\cc_m}$ is the symmetric matrix (\ref{SYMMETRIC}) evaluated at the $m$th approximation.
The  projector $\PP_{m}$ is not obligatory thermodynamic at that step, and it is taken as follows:
\begin{equation}
\label{proj_nonthermo} \PP_m\xx=\sum_{i=1}^{k}\frac{\partial\cc_m}{\partial c_i} (\mm_i,\xx)+
\sum_{i=1}^{l}\frac{\partial\cc_m}{\partial B_i} (\bb_i,\xx).
\end{equation}
(iii). {\it Dynamics.} Dynamics on the $m$th manifold is obtained with the thermodynamic
re-parameterization.

In the next section we shall illustrate how this all works.

\section{Illustration: Two-step catalytic reaction}
\label{EX}

Here we consider a two-step four-component reaction with one catalyst $A_2$:
\begin{equation}
\label{mech_ex} A_1+A_2\rightleftharpoons\ A_3\rightleftharpoons A_2 + A_4.
\end{equation}
We assume the Lyapunov function of the form (\ref{gfun}), $G=\sum_{i=1}^{4}c_i[\ln(c_i/c_i^{\rm
eq})-1]$. The kinetic equation for the four--component vector of concentrations,
$\cc=(c_1,c_2,c_3,c_4)$, has the form
\begin{equation}
\label{1exequ} \dot{\cc}=\bgamma_1W_1+\bgamma_2W_2.
\end{equation}
Here $\bgamma_{1,2}$ are stoichiometric vectors,
\begin{equation}
\label{stochio_ex} \bgamma_1=(-1,-1,1,0),\  \bgamma_2=(0,1,-1,1),
\end{equation}
while functions $W_{1,2}$ are reaction rates:
\begin{equation}
\label{rates_ex} W_1=k^+_1c_1c_2-k^-_1c_3,\ W_2=k^+_2c_3- k^-_2c_2 c_4.
\end{equation}
Here $k_{1,2}^{\pm}$ are reaction rate constants. The system under consideration has two conservation
laws,
\begin{equation}
\label{cons_ex} c_1+c_3+c_4=B_1,\ c_2 +c_3=B_2,
\end{equation}
or $(\bb_{1,2},\cc)=B_{1,2}$, where $\bb_1=(1,0,1,1)$ and $\bb_2=(0,1,1,0)$. The nonlinear system
(\ref{1exequ}) is effectively two-dimensional, and we consider a one-dimensional reduced description.

We have chosen the concentration of the specie $A_1$ as the variable of reduced description: $M=c_1$,
and $c_1=(\mm,\cc)$, where $\mm=(1,0,0,0)$. The initial manifold $\cc_0(M)$ was taken as the
quasi-equilibrium approximation, i.e. the vector function $\cc_0$ is the solution to the problem:
\begin{equation}
\label{qe} G\to{\rm min}\ {\rm for}\ (\mm,\cc)=c_1,\ (\bb_1,\cc)=B_1,\ (\bb_2,\cc)=B_2.
\end{equation}
The solution to the problem (\ref{qe}) reads:
\begin{eqnarray}
\label{1ex1} c_{01}&=&c_1,\\\nonumber c_{02}&=&B_2-\phi(c_1), \\\nonumber
c_{03}&=&\phi(c_1),\\\nonumber c_{04}&=&B_1-c_1-\phi(c_1),\\\nonumber
\phi(M)&=&A(c_1)-\sqrt{A^2(c_1)-B_2(B_1-c_1)},\\\nonumber A(c_1)&=&\frac{B_2(B_1-c^{\rm
eq}_1)+c_3^{\rm eq}(c_1^{\rm eq}+ c_3^{\rm eq}-c_1)}{2c_3^{\rm eq}} .
\end{eqnarray}

\begin{figure}[p]
\centering{
 \caption{Images of the initial quasi-equilibrium manifold (bold line)
 and the first two corrections
(solid normal lines) in the phase plane $[c_1,c_3]$ for two-step catalytic reaction. Dashed lines are
individual trajectories.} \label{Fig1}}
\end{figure}

The thermodynamic projector associated with the manifold (\ref{1ex1}) reads:
\begin{equation}
\label{proj_0} \PP_0\xx=\frac{\partial\cc_0}{\partial c_1}(\mm,\xx) +\frac{\partial\cc_0}{\partial
B_1}(\bb_1,\xx)+ \frac{\partial\cc_0}{\partial B_2}(\bb_2,\xx).
\end{equation}
Computing $\bDelta_0=[1-\PP_0]\JJ(\cc_0)$ we find that the inequality (\ref{DEFECT}) takes place, and
thus the manifold $\cc_0$ is not invariant. The first correction, $\cc_1=\cc_0+\delta\cc$, is found
from the linear algebraic system (\ref{corr_eq})
\begin{eqnarray}
\label{LIN_SYSTEM} (1-\PP_0)\LL'_{0}\delta\cc&=&-[1-\PP_0]\JJ(\cc_0),\\\nonumber \delta c_1&=&0
\nonumber \\ \delta c_1+\delta c_3+\delta c_4&=&0 \nonumber \\ \delta c_3+\delta c_2&=&0,
\label{correction1}
\end{eqnarray}
where the symmetric $4\times 4$ matrix $\LL'_{0}$ has the form (we write $0$ instead of $\cc_0$ in
the subscript in order to simplify notations):
\begin{equation}
\label{SL} L'_{0, kl}=-\gamma_{1k} \frac{W_1^+(\cc_0)+W_1^-(\cc_0)}{2} \frac{\gamma_{1l}}{c_{0l}}
-
\gamma_{2k} \frac{W_2^+(\cc_0)+W_2^-(\cc_0)}{2} \frac{\gamma_{2l}}{c_{0l}}
\end{equation}
The explicit solution $\cc_1(c_1,B_1,B_2)$ to the linear system (\ref{LIN_SYSTEM}) is easily found,
and we do not reproduce it here. The process was iterated. On the $k+1$ iteration, the following
projector $\PP_{k}$ was used:
\begin{equation}
\label{proj_k} \PP_k\xx=\frac{\partial\cc_k}{\partial c_1}(\mm,\xx)+ \frac{\partial\cc_k}{\partial
B_1}(\bb_1,\xx)+ \frac{\partial\cc_k}{\partial B_2}(\bb_2,\xx).
\end{equation}
Notice that projector $\PP_k$ (\ref{proj_k}) is the thermodynamic projector only if $k=0$. As we have
already mentioned it above, in the process of finding the corrections to the manifold, the
non-thermodynamic projectors are allowed. The linear equation at the $k+1$ iteration is thus obtained
by replacing $\cc_0$, $\PP_0$, and $\LL'_0$ with $\cc_k$, $\PP_k$, and $\LL'_k$ in all the entries of
the Eqs.\ (\ref{LIN_SYSTEM}) and (\ref{SL}).

Once the manifold $\cc_k$ was obtained on the $k$th iteration, we derived the corresponding dynamics
by introducing the thermodynamic parameterization (and the corresponding thermodynamic projector)
with the help of the function (\ref{param_thermo}). The resulting dynamic equation for the variable
$c_1$ in the $k$th approximation has the form:
\begin{equation}
\label{1ex3} (\bnabla G\bigm|_{\cc_k},\partial\cc_k/\partial c_1)\dot{c_1}= (\bnabla
G\bigm|_{\cc_k},\JJ(\cc_k)).
\end{equation}
Here $[\bnabla G\bigm|_{\cc_k}]_i=\ln[c_{ki}/ c^{\rm eq}_i]$.

Analytic results were compared with the results of the numerical integration. The following set of
parameters was used:
\begin{eqnarray*}
k^+_1=1.0,\ k^-_1=0.5,\  k^+_2=0.4 ,\ k^-_2=1.0;\\ c_1^{\rm eq}=0.5,\ c_2^{\rm eq}=0.1,\ c_3^{\rm
eq}=0.1,\ c_4^{\rm eq}=0.4,\\ B_1=1.0    ,\ \ B_2=0.2   .
\end{eqnarray*}
Direct numerical integration of the system has demonstrated that the manifold $c_3= c^{\rm eq}_3$ in
the plane $(c_1,c_3)$ attracts all individual trajectories. Thus, the reduced description in this
example should extract this manifold.

Fig.\ \ref{Fig1}  demonstrates the quasi-equilibrium manifold (\ref{1ex1}) and the first two
corrections found analytically. It should be stressed that we spend no special effort on the
construction of the initial approximation, that is, of the quasi-equilibrium manifold, have not used
any information about the Jacobian field (unlike, for example, the ILDM or CSP methods discussed
above) etc. It is therefore not surprising that in this way chosen  initial quasi-equilibrium
approximation is in a rather  poor agreement with the reduced description. However, it should be
appreciated that the further corrections {\it rapidly} improve the situation while no small parameter
considerations were used. This confirms our expectation of the advantage of using the  iteration
methods in comparison to methods based on a small parameter expansions for model reduction problems.

\section{Relaxation methods} \label{relax}

Relaxation method is an alternative to the Newton iteration method described in section \ref{DC}. It
is a one-dimensional Galerkin approximation for the linearized invariance equation
(\ref{METHOD},\ref{uni_con}). We shall solve the invariance equation (\ref{METHOD},\ref{uni_con}) (or
symmetric invariance equation (\ref{corr_eq},\ref{corr_cond}) in projection on the defect of
invariance (\ref{DEFECT}) $\bDelta= [1-\PP_{\cc}]\JJ(\cc).$

Let $\bOmega_0$ be the initial approximation to the invariant manifold, and we seek the first
correction, $ \cc_1=\cc_0+\tau_1(\cc_0)\bDelta(\cc_0),$ where function $\tau(\cc_0)$ has a dimension
of the time, and is found from the condition that the linearized vector field attached to the points
of the new manifold is orthogonal to the initial defect,
\begin{equation}\label{terminator1}
  \langle\bDelta(\cc_0),(1-\PP_{\cc_0})[\JJ(\cc_0)+
  \tau_1(\cc_0)(D_{\cc}\JJ)_{\cc_0}\bDelta_{\cc_0}]\rangle_{\cc_0}=0.
\end{equation}
Explicitly,
\begin{equation}\label{terminator2}
 \tau_1(\cc_0)=-\frac{\langle\bDelta_{\cc_0},\bDelta_{\cc_0}\rangle_{\cc_0}}
 {\langle\bDelta_{\cc_0},(D_{\cc}\JJ)_{\cc_0}\bDelta_{\cc_0}\rangle_{\cc_0}}.
\end{equation}
Further steps $\tau_k(\cc)$ are found in the same way. It is clear from the latter equations that the
step of the relaxation method is equivalent to the Galerkin approximation for solving the step of the
Newton method. Actually, the relaxation method was first introduced in these terms in
\cite{KZFPLANL}. An advantage of equation (\ref{terminator2}) is the explicit form of the size of the
steps $\tau_k(\cc)$. This method was successfully applied to the Fokker-Plank equation
\cite{KZFPLANL}.

\section{Method of invariant manifold without a priori parameterization}
\label{PARAMETERIZATION}

Formally, the method of invariant manifold does not require a global parameterization of the
manifolds. However, in most of the cases, one makes use of a priori defined ``macroscopic'' variables
$M$. This is motivated by the choice of quasi-equilibrium initial approximations.

Let a manifold $\bOmega$ be defined in the phase space of the system, its tangent space in the point
$\cc$  be $T_{\cc}\bOmega$. How to define the projector of the whole concentrations space onto
$T_{\cc}\bOmega$ without using any a priori parameterization of $\bOmega$?

The basis of the answer to this question is the condition of thermodynamicity (\ref{therm1}). Let us
denote $E$ as the concentration space, and consider the problem of the choice of the projector in the
quadratic approximation to the thermodynamic potential $G$:

\begin{equation}
G_{\rm q}=(\bg,\HH_{\cc}\Delta\cc) +\frac{1}{2}(\Delta\cc,\HH_{\cc}\Delta\cc)
=\langle\bg,\Delta\cc\rangle + \frac{1}{2}\langle\Delta\cc,\Delta\cc\rangle, \label{Gquadratic}
\end{equation}
where $\HH_{\cc}$ is the matrix of the second-order derivatives of $G$ (\ref{MATRIX}),
$\bg=\HH^{-1}_{\cc}\bnabla G$, $\Delta\cc$ is the deviation of the concentration vector from the
expansion point.

Let a linear subspace $T$ be given in the concentrations space $E$. {\it Problem:} For every
$\Delta\cc+T$, and for every $\bg\in E$, define  a subspace $L_{\Delta\cc}$ such that: (i)
$L_{\Delta\cc}$ is a complement of $T$ in $E$:
\[L_{\Delta\cc}+T=E,\ L_{\Delta\cc}\cap T=\{\bZERO\}.\]
(ii) $\Delta\cc$ is the point of minimum of $G_{\rm q}$ on $L_{\Delta\cc}+\Delta\cc$:
\begin{equation}
\label{minQ} \Delta\cc=\arg\min_{\xx-\Delta\cc\in L_{\Delta\cc}}G_{\rm q}(\xx).
\end{equation}
Besides (i) and (ii), we also impose the requirement of a {\it maximal smoothness} (analyticity) on
$L_{\Delta\cc}$ as a function of $\bg$ and $\Delta\cc$. Requirement (\ref{minQ}) implies that
$\Delta\cc$ is the quasi-equilibrium point for the given $L_{\Delta\cc}$, while the problem in a
whole is the {\it inverse} quasi-equilibrium problem: We construct $L_{\Delta\cc}$ such that $T$ will
be the quasi-equilibrium manifold. Then subspaces $L_{\Delta\cc}$ will actually be the kernels of the
quasi-equilibrium projector.

Let $\ff_1,\dots,\ff_k$ be the orthonormalized with respect to $\langle\cdot,\cdot\rangle$ scalar
product basis of $T$, vector $\hh$ be orthogonal to $T$, $\langle\hh,\hh\rangle=1$,
$\bg=\alpha\ff_1+\beta\hh$. Condition (\ref{minQ}) implies that the vector $\bnabla G$ is orthogonal
to $L_{\Delta\cc}$ in the point $\Delta\cc$.

Let us first consider the case $\beta=0$. The requirement of analyticity of $L_{\Delta\cc}$ as the
function of $\alpha$ and $\Delta\cc$ implies $L_{\Delta\cc}=L_{\bZERO}+o(1)$, where
$L_{\bZERO}=T^{\perp}$ is the orthogonal completement of $T$ with respect to scalar product
$\langle\cdot,\cdot\rangle$. The constant solution, $L_{\Delta\cc}\equiv L_{\bZERO}$ also satisfies
(\ref{minQ}). Let us fix $\alpha\ne0$, and extend this latter solution to $\beta\ne0$. With this, we
obtain a basis, $\bl_1,\dots,\bl_{n-k}$. Here is the simplest construction of this basis:
\begin{equation}
\label{ort1} \bl_1=\frac{\beta\ff_1-(\alpha+\Delta c_1)\hh}{(\beta^2+(\alpha+\Delta c_1)^2)^{1/2}},
\end{equation}
where $\Delta c_1=\langle\Delta\cc,\ff_1\rangle$ is the first component in the expansion,
$\Delta\cc=\sum_i\Delta c_i\ff_i$. The rest of the basis elements, $\bl_2,\dots,\bl_{n-k}$ form the
orthogonal completement of $T\oplus(\hh)$ with respect to scalar product $\langle\cdot,\cdot\rangle$,
$(\hh)$ is the line spanned by $\hh$.

Dependence $L_{\Delta\cc}$ (\ref{ort1}) on $\Delta\cc$, $\alpha$ and $\beta$ is singular: At
$\alpha+\Delta c_1$, vector $\bl_1\in T$, and then  $L_{\Delta\cc}$ is not the completement of $T$ in
$E$ anymore. For $\alpha\ne0$, dependence $L_{\Delta\cc}$ gives one of the solutions to the inverse
quasi-equilibrium problem in the neighborhood of zero in $T$. We are interested only in the limit,
\begin{equation}
\label{lim1} \lim_{\Delta\cc\to\bZERO}L_{\Delta\cc}={\rm Lin}\left\{
\frac{\beta\ff_1-\alpha\hh}{\sqrt{\alpha^2+\beta^2}},\bl_2,\dots,\bl_{n-k}\right\}.
\end{equation}

Finally, let us define now the projector $\PP_{\cc}$ of the space $E$ onto $T_{\cc}\bOmega$. If
$\HH^{-1}_{\cc}\bnabla G\in T_{\cc}\bOmega$, then $\PP_{\cc}$ is the orthogonal projector with
respect to the scalar product
 $\langle\cdot,\cdot\rangle$:

\begin{equation}
\PP_{\cc}\zz=\sum_{i=1}^k\ff_i\langle \ff_i,\zz\rangle.
\end{equation}
If $\HH^{-1}_{\cc}\bnabla G \notin T_{\cc}\bOmega$, then, according to Eq.\ (\ref{lim1}),
\begin{equation}
\label{result_projector} \PP_{\cc}\zz=\frac{\langle \ff_1,\zz\rangle -\langle \bl_1,\zz\rangle
\langle \ff_1,\bl_1\rangle} {1-\langle \ff_1,\bl_1\rangle^2}\ff_1+ \sum_{i=2}^k\ff_i\langle
\ff_i,\zz\rangle,
\end{equation}
where $\{\ff_1,\dots,\ff_k\}$ is the orthonormal with respect to $\langle\cdot,\cdot \rangle$ basis
of $T_{\cc}\bOmega$, $\hh$ is orthogonal to $T$, $\langle \hh,\hh\rangle=1$, $\HH^{-1}_{\cc}\bnabla
G=\alpha\ff_1+\beta\hh$, $\bl_1=(\beta\ff_1-\alpha\hh)/\sqrt{\alpha^2+\beta^2}$, $\langle
\ff_1,\bl_1\rangle=\beta/\sqrt{\alpha^2+\beta^2}$.

Thus, for solving the invariance equation iteratively, one needs only projector $\PP_{\cc}$
(\ref{result_projector}), and one does not need a priori parameterization of $\bOmega$ anymore.

\section{Method of invariant grids} \label{grid}

Elsewhere above in this paper, we considered the invariant manifold, and methods for their
construction, without addressing the question of how to implement it in a {\it constructive way}. In
most of the works (of us and of other people on similar problems), analytic forms were needed to
represent manifolds. However, in order to construct manifolds of a relatively low dimension,
grid-based representations of manifolds become a relevant option. The \textit{Method of invariant
grids} (MIG) was suggested recently in \cite{InChLANL}.

The principal idea of (MIG) is to find a mapping of finite-dimensional grids into the phase space of
a dynamic system. That is we construct not just a point approximation of the invariant manifold, but
an {\it invariant grid}. When refined, in the limit it is expected to converge, of course, to the
invariant manifold, but it is a separate, independently  defined object.

Let's denote $L=\RR^n$, $\mbox{\bf G}$ is a discrete subset of $\RR^n$. A natural choice would be a
regular grid, but, this is not crucial from the point of view of the general formalism. For every
point $y \in \mbox{\bf G}$, a neighborhood of $y$ is defined: $V_y \subset \mbox{\bf G}$, where $V_y$
is a finite set, and, in particular, $y \in V_y$. On regular grids, $V_y$ includes, as a rule, the
nearest neighbors of $y$. It may also include next to nearest points.

For our purposes, one should define a grid differential operator. For every function, defined on the
grid, also all derivatives are defined:

\begin{equation} \label{diffgrid}
\left.\frac{\partial f}{\partial y_i}\right|_{y\in \mbox{\bf G}} = \sum_{z \in V_y} q_i(z,y)f(z),
i=1,\ldots n.
\end{equation}

\noindent where $q_i(z,y)$ are some coefficients.

Here we do not specify the choice of the functions $q_i(z,y)$. We just mention in passing that, as a
rule, equation (\ref{diffgrid}) is established using some interpolation of $f$ in the neighborhood of
$y$ in $\RR^n$ by some differentiable functions (for example, polynomial). This interpolation is
based on the values of $f$ in the points of $V_y$. For regular grids, $q_i(z,y)$ are functions of the
difference $z-y$. For some $y$s which are close to the edges of the grid, functions are defined only
on the part of $V_y$. In this case, the coefficients in (\ref{diffgrid}) should be modified
appropriately in order to provide an approximation using available values of $f$. Below we will
assume this modification is always done. We also assume that the number of points in the neighborhood
$V_y$ is always sufficient to make the approximation possible. This assumption restricts the choice
of the grids $\mbox{\bf G}$. Let's call {\it admissible} all such subsets $\mbox{\bf G}$, on which
one can define differentiation operator in every point.

Let $F$ be a given mapping of some admissible subset $\mbox{\bf G} \subset \RR^n$ into $U$. For every
$y \in V$ we define tangent vectors:

\begin{equation}\label{tanspace}
T_y = Lin\{\bg_i\}_1^n,
\end{equation}

\noindent where vectors $\bg_i (i=1, \ldots n)$ are partial derivatives (\ref{diffgrid}) of the
vector-function $F$:

\begin{equation}\label{bastan}
\bg_i = \frac{\partial F}{\partial y_i} = \sum_{z \in V_y} q_i(z,y)F(z),
\end{equation}

or in the coordinate form:

\begin{equation}\label{bastanco}
(\bg_i)_j = \frac{\partial F_j}{\partial y_i} = \sum_{z \in V_y} q_i(z,y)F_j(z).
\end{equation}

\noindent Here $(\bg_i)_j$ is the $j$th coordinate of the vector $(\bg_i)$, and $F_j(z)$ is the $j$th
coordinate of the point $F(z)$.

The grid $\mbox{\bf G}$ is {\it invariant}, if for every node $y \in \mbox{\bf G}$ the vector field
$J(F(y))$ belongs to the tangent space $T_y$ (here $J$ is the right hand site of the kinetic
equations (\ref{reaction})).

So, the definition of the invariant grid includes:

\noindent 1) Finite admissible subset $\mbox{\bf G} \subset \RR^n$;

\noindent 2) A mapping $F$ of this admissible subset $\mbox{\bf G}$ into $U$ (where $U$ is the phase
space for kinetic equations (\ref{reaction}));

\noindent 3) The differentiation formulas (\ref{diffgrid}) with given coefficients $q_i(z,y)$;

The grid invariance equation has a form of inclusion:

$$\JJ(F(y)) \in T_y \: \mbox{for every} \: y \in \mbox{\bf G},$$

\noindent or a form of equation:

$$ (1-\PP_{F(y)})\JJ(F(y))=0 \: \mbox{for every} \: y \in \mbox{\bf G},$$

\noindent where $\PP_{F(y)}$ is the thermodynamic projector (\ref{result_projector}).

The grid differentiation formulas (\ref{diffgrid}) are needed, in the first place, to establish the
tangent space $T_y$, and the null space of the thermodynamic projector $\PP_{F(y)}$ in each node. It
is important to realise that locality of construction of thermodynamic projector enables this without
a need for a global parametrization.

Basically, in our approach, the grid specifics are in: (a) differentiation formulas, (b) grid
construction strategy (the grid can be extended, contracted, refined, etc.) The invariance equations
(\ref{DEFECT}), the iteration Newton method (\ref{METHOD},\ref{uni_con}), and the formulas of the
relaxation approximation (\ref{terminator2}) do not change at all. For convenience, let us repeat all
these formulas in the grid context.

Let $\cc=F(y)$ be position of a grid's node $y$ immersed into phase space $U$. We have set of tangent
vectors $\bg_i(x)$, defined in $\cc$ (\ref{bastan}), (\ref{bastanco}). Thus, the tangent space $T_y$
is defined by (\ref{tanspace}). Also, one has the thermodynamic Lyapunov function $G(\cc)$, the
linear functional $D_{\cc}G|_{\cc}$, and the subspace $T_{0y}=T_y\bigcap\ker D_{\cc}G|_{\cc}$ in
$T_y$. Let $T_{0y}\neq T_y$. In this case we have a vector ${\ee}_y\in T_y$, orthogonal to $T_{0y}$,
$D_{\cc}G|_{\cc}({\ee}_y)=1$. Then, the thermodynamic projector is defined as:

\begin{equation}\label{projgengrid}
  \PP_{\cc}\bullet=\PP_{0\cc}\bullet+{\ee}_y D_{\cc}G|_{\cc}\bullet,
\end{equation}

\noindent where $\PP_{0\cc}$ is the orthogonal projector on $T_{0y}$ with respect to the entropic
scalar product $\langle , \rangle_x$.

If $T_{0y} = T_y$, then the thermodynamic projector is the orthogonal projector on $T_y$ with respect
to the entropic scalar product $\langle , \rangle_{\cc}$.

For the Newton method with incomplete linearization, the equations for calculating new node position
$\cc '=\cc+\delta \cc$ are:

\begin{equation}\label{Nm1grid}
 \left\{\begin{array}{l}
   \PP_{\cc}\delta \cc=0 \\
   (1-\PP_{\cc})(\JJ(\cc)+D\JJ(\cc)\delta \cc)=0.
 \end{array}\right.
\end{equation}

\noindent Here $D\JJ(\cc)$ is a matrix of derivatives of $\JJ$, calculated in $\cc$. The self-adjoint
linearization may be useful too (see section \ref{SA}).

Equation (\ref{Nm1grid}) is a system of linear algebraic equations. In practice, it is convenient to
choose some orthonormal (with respect to the entropic scalar product) basis $\bb_i$ in $ker
\PP_{\cc}$. Let $r=dim(ker \PP_{\cc} )$. Then $\delta \cc = \sum_{i=1}^r{\delta_i \bb_i}$, and the
system looks like

\begin{equation}\label{Nm1grid1}
   \sum_{k=1}^{r}{\delta_k \langle {\bf \bb_i} , D\JJ(\cc){\bb_k} \rangle _{\cc}
   = -  \langle \JJ(\cc) , \bb_i \rangle_{\cc} }, i = 1...r.
\end{equation}

Here $\langle ,  \rangle_{\cc}$ is the entropic scalar product (\ref{ESP}). This is the system of
linear equations for adjusting the node position accordingly to the Newton method with incomplete
linearization.

For the relaxation method, one needs to calculate the defect $\Delta_{\cc} = (1-\PP_{\cc})\JJ(\cc)$,
and the relaxation step

\begin{equation}\label{terminator2grid}
 \tau(x)=-\frac{\langle\Delta_{\cc} , \Delta_{\cc}\rangle_{\cc}}
 {\langle\Delta_{\cc} , D\JJ(\cc)\Delta_{\cc}\rangle_{\cc}}.
\end{equation}

Then, new node position $x'$ is calculated as

\begin{equation}\label{relaxgrid}
\cc' = \cc+\tau(\cc)\Delta_{\cc}.
\end{equation}

This is the equation for adjusting the node position according to the relaxation method.

\subsection{Grid construction strategy}

From all reasonable strategies of the invariant grid construction we will consider here the following
two: {\it growing lump} and {\it invariant flag}.

\subsubsection{Growing lump}

In this strategy one chooses as initial the equilibrium point $y^*$. The first approximation is
constructed as $F(y^*)=\cc^*$, and for some initial $V_0$ ($V_{y^*} \subset V_0$) one has
$F(y)=\cc^*+A(y-y^*)$, where $A$ is an isometric embedding (in the standard Euclidean metrics) of
$\RR^n$ in $E$.

For this initial grid one makes a fixed number of iterations of one of the methods chosen (Newton's
method with incomplete linearization or the relaxation method), and, after that, puts
$V_1=\bigcup_{y\in V_0}V_y$ and extends $F$ from $V_0$ onto $V_1$ using linear extrapolation and the
process continues. One of the possible variants of this procedure is to extend the grid from $V_i$ to
$V_{i+1}$ not after a fixed number of iterations, but when  the invariance defect $\Delta_y$ becomes
smaller than a given $\epsilon$ (in a given norm, which is entropic, as a rule), for all nodes $y\in
V_i$. The lump stops growing when it reaches the boundary and is within a given accuracy
$\|\Delta\|<\epsilon$.

\subsubsection{Invariant flag}

For the invariant flag one uses sufficiently regular grids $\mbox{\bf G}$, in which many points are
situated on the coordinate lines, planes, etc. One considers the standard flag $\RR^0 \subset \RR^1
\subset \RR^2 \subset ... \subset \RR^n$ (every next space is constructed by adding one more
coordinate). It corresponds to a succession of grids $\{y\} \subset \mbox{\bf G}^1 \subset \mbox{\bf
G}^2 ... \subset \mbox{\bf G}^n$ , where $\{y^*\} = \RR^0$, and $\mbox{\bf G}^i$ is a grid in
$\RR^i$.

First, $y^*$ is mapped in $\cc^*$ and further $F(y^*)=\cc^*$. Then an invariant grid is constructed
on $V^1\subset \mbox{\bf G}^1$ (up to the boundaries $U$ and within a given accuracy
$\|\Delta\|<\epsilon$). After the neighborhoods in $\mbox{\bf G}^2$ are added to the points $V^1$,
and, using such extensions, the grid $V^2\subset \mbox{\bf G}^2$ is constructed (up to the boundaries
and within a given accuracy) and so on, until $V^n\subset \mbox{\bf G}^n$ will be constructed.

We must underline here that, constructing the k-th grid $V^k\subset \mbox{\bf G}^k$, the important
role of the grids of smaller dimension $V^0\subset ... \subset V^{k-1} \subset V^k$ embedded in it,
is preserved. The point $F(y^*)=x^*$ is preserved. For every $y\in V^q$ ($q<k$) the tangent vectors
$g_1,...,g_q$ are constructed, using the differentiation operators (\ref{diffgrid}) on the whole
$V^k$. Using the tangent space $T_y = Lin\{g_1,..,g_q\}$, the projector $\PP_{F(y)}$ is constructed,
the iterations are applied and so on. All this is done to obtain a succession of embedded invariant
grids, given by the same map $F$.

\subsubsection{Boundaries check and the entropy}

We construct grid mapping of $F$ onto the finite set $V\in \mbox{\bf G}$. The technique of checking
if the grid still belongs to the phase space $U$ of kinetic system  ($F(V)\subset U$) is quite
straightforward: all the points $y\in V$ are checked to belong to $U$. If at the next iteration a
point $F(y)$ leaves $U$, then it is returned inside by a homothety transform with the center in
$x^*$. Since the thermodynamic Lyapunov function is a convex function, the homothety contraction with
the center in $x^*$ decreases it monotonously. Another variant is cutting off the points leaving $U$.

By the way it was constructed (\ref{result_projector}) the kernel of the thermodynamic projector is
annulled by the entropy differential. Thus, in the first order, steps in the Newton method with
incomplete linearization (\ref{METHOD},\ref{uni_con}) as well as in the relaxation methods
(\ref{terminator1}),(\ref{terminator2}) do not change the entropy. But, if the steps are quite large,
then the increasing of the thermodynamic Lyapunov function can become essential and the points are
returned on their level by the homothety contraction with the center in the equilibrium point.

\subsection{Instability of fine grids}

\begin{figure}
\centering{
\caption{Grid instability. For small grid steps approximations in the calculation of grid derivatives
lead to the grid instability effect. On the figure several successive iterations of the algorithm
without adaptation of the time step are shown that lead to undesirable ``oscillations", which
eventually destruct the grid starting from one of it's ends. \label{diver}}}
\end{figure}

When one reduces the grid step (spacing between the nodes) in order to get a finer grid, then,
starting from a definite step, it is possible to face the problem of the Courant instability
\cite{Cour}. Instead of converging, at the every iteration the grid becomes entangled (see
Fig.~\ref{diver}).

The way to get rid off this instability is well-known. This is decreasing the time step. Instead of
the real time step, we have a shift in the Newtonian direction. Formally, we can assign for one
complete step in the Newtonian direction a value $h=1$. Let us consider now the Newton method with an
arbitrary $h$. For this, let us find $ \delta \cc =\delta F(y)$ from (\ref{Nm1grid}), but we will
change $\delta \cc$ proportionally to $h$: the new value of $\cc_{n+1}=F_{n+1}(y)$ will be equal to

\begin{equation}
F_{n+1}(y) = F_n(y)+h_n\delta F_n(y)
\end{equation}

\noindent where the lower index $n$ denotes the step number.

One way to choose the $h$ step value is to make it adaptive, controlling the average value of the
invariance defect $\|\Delta_y\|$ at every step. Another way is the convergence control: then $\sum
h_n$ plays a role of time.

Elimination of Courant instability for the relaxation method can be made quite analogously.
Everywhere the step $h$ is maintained as big as it is possible without convergence problems.

\subsection{What space is the most appropriate for the grid construction?}

For the kinetics systems there are two distinguished representations of the phase space:

\begin{itemize}
\item The densities space (concentrations, energy or probability densities, etc.)
\item The spaces of conjugate intensive quantities, potentials (temperature, chemical potentials, etc.)
\end{itemize}

The density space is convenient for the construction of quasi-chemical representations. Here the
balance relations are linear and the restrictions are in the form of linear inequalities (the
densities themselves or some linear combinations of them must be positive).

The conjugate variables space is convenient in the sense that the equilibrium conditions, given the
linear restrictions on the densities, are in the linear form (with respect to the conjugate
variables). In these spaces the quasi-equilibrium manifolds exist in the form of linear subspaces
and, vise versa, linear balance equations turns out to be equations of the conditional entropy
maximum.

The duality we've just described is very well-known and studied in details in many works on
thermodynamics and Legendre transformations \cite{Leg1,Leg2}. In the previous section, the grids were
constructed in the density space. But the procedure of constructing them in the space of the
conjugate variables seems to be more consistent. The principal argument for this is the specific role
of quasi-equilibrium, which exists as a linear manifold. Therefore, linear extrapolation gives a
thermodynamically justified quasi-equilibrium approximation. Linear approximation of the slow
invariant manifold in the neighborhood of the equilibrium in the conjugate variables space already
gives the global quasi-equilibrium manifold, which corresponds to the motion separation (for slow and
fast motions) in the neighborhood of the equilibrium point.

For the mass action law, transition to the conjugate variables is simply the logarithmic
transformation of the coordinates.

\subsection{Carleman's formulas in the analytical invariant manifolds approximations. First profit
from analyticity: superresolution}

When constructing invariant grids, one must define the differential operators (\ref{diffgrid}) for
every grid node. For calculating the differential operators in some point $y$, an interpolation
procedure in the neighborhood of $y$ is used. As a rule, it is an interpolation by a low-order
polynomial, which is constructed using the function values in the nodes belonging to the
neighbourhood of $y$ in $\mbox{\bf G}$. This approximation (using values in the closest nodes) is
natural for smooth functions. But, for the systems (\ref{reaction}) with analytical right hand side
we are looking for the {\it analytical} invariant manifold (due to Lyapunov auxiliary theorem
\cite{Lya,Kazantzis00}). Analytical functions have much more ``rigid" structure than the smooth ones.
One can change a smooth function in the neighborhood of any point in such a way, that outside this
neighborhood the function will not change. In general, this is not possible for analytical functions:
a kind of ``long-range" effect takes place (as is well known) .

The idea is to use this effect and to reconstruct some analytical function $f_{\mbox{\scriptsize \bf
G}}$ using function given on $\mbox{\bf G}$. There is one important requirement: if these values on
$\mbox{\bf G}$ are values (given in the points of $\mbox{\bf G}$) of some function $f$ which is
analytical in the given neighborhood $U$, then if the $\mbox{\bf G}$ is refined ``correctly", one
must have $f_{\mbox{\scriptsize \bf G}} \rightarrow f$. The sequence of reconstructed function
$f_{\mbox{\scriptsize \bf G}}$ should should converge to the ``proper" function $f$.

What is the ``correct refinement"? For smooth functions for the convergence $f_{\mbox{\scriptsize \bf
G}} \rightarrow f$  it is necessary and sufficient that, in the course of refinement, $\mbox{\bf G}$
would approximate the whole $U$ with arbitrary accuracy. For analytical functions it is only
necessary that, under the refinement, $\mbox{\bf G}$ would approximate some uniqueness set
\footnote{Let's remind to the reader that $A \subset U$ is called {\it uniqueness set} in $U$ if for
analytical in $U$ functions $\psi$ and $\varphi$ from $\psi|_A \equiv \varphi|_A$ it follows
$\psi=\varphi$.} $A \subset U$. Suppose we have a sequence of grids $\mbox{\bf G}$, each next is
finer than previous, which approximates a set $A$. For smooth functions, using function values
defined on the grids, one can reconstruct the function in $A$. For analytical functions, if the
analyticity area $U$ is known, and $A$ is a uniqueness set in $U$, then one can reconstruct the
function in $U$. The set $U$ can be essentially bigger than $A$; because of this such extension was
named as {\it superresolution effects} \cite{Aiz,GR99}. There exist constructive formulas for
construction of analytical functions $f_{\mbox{\scriptsize \bf G}}$ for different areas $U$,
uniqueness sets $A\subset U$ and for different ways of discrete approximation of $A$ by a sequence of
fined grids $\mbox{\bf G}$ \cite{Aiz}. Here we provide only one Carleman's formula which is the most
appropriate for our purposes.

Let area $U=Q^{n}_\sigma \subset \CC^n$ be a product of strips $Q_\sigma\subset C$,
$Q_\sigma=\{z|\mbox{Im} z<\sigma\}$. We will construct functions holomorphic in $Q^{n}_\sigma$. This
is effectively equivalent to the construction of real analytical functions $f$ in whole $\RR^n$ with
a condition on the convergence radius $r(x)$ of the Taylor series for $f$ as a function of each
coordinate: $r(x) \geq \sigma$ in every point $x \in \RR^n$.

The sequence of fined grids is constructed as follows: let for every $l=1,...,n$ a finite sequence of
distinct points $N_l\subset D_\sigma$ be defined:

\begin{equation}
N_l = \{x_{lj}|j = 1,2,3 ... \}, x_{lj} \neq x_{li} \hspace{5pt} for \hspace{5pt} i\neq j
\end{equation}

The uniqueness set $A$, which is approximated by a sequence of fined finite grids, has the form:

\begin{equation}
A = N_1 \times N_2 \times ... \times N_n = \{(x_{1i_1},x_{2i_2},..,x_{ni_n})|i_{1,..,n}=1,2,3,...\}
\end{equation}

The grid $\mbox{\bf G}_m$ is defined as the product of initial fragments $N_l$ of length $m$:

\begin{equation}
\mbox{\bf G}_m = \{(x_{1i_1},x_{2i_2}...x_{ni_n})|1\leq i_{1,..,n}\leq m\}
\end{equation}

Let's denote $\lambda=2\sigma/\pi$ ($\sigma$ is a half-width of the strip $Q_\sigma$). The key role
in the construction of the Carleman's formula is played by the functional $\omega_m^\lambda(u,p,l)$
of 3 variables: $u \in U = Q^n_\sigma$, $p$  is an integer, $1\leq p \leq m$, $l$ is an integer,
$1\leq p \leq n$. Further $u$ will be the coordinate value in the point where the extrapolation is
calculated, $l$ will be the coordinate number, and $p$  will be an element of multi-index
$\{i_1,...,i_n\}$ for the point $(x_{1i_1},x_{2i_2},...,x_{ni_n})\in \mbox{\bf G}$:

\begin{equation}\label{Carleman}
\omega_m^\lambda(u,p,l) = \frac{(e^{\lambda x_{lp}}+e^{\lambda \bar{x}_{lp}})(e^{\lambda
u}-e^{\lambda x_{lp}})} {\lambda(e^{\lambda u}+e^{\lambda \bar{x}_{lp}})(u-x_{lp})e^{\lambda
x_{lp}}}\times \prod_{j=1 j\neq p}^m{\frac{(e^{\lambda x_{lp}}+e^{\lambda \bar{x}_{lj}})(e^{\lambda
u}-e^{\lambda x_{lj}})}{(e^{\lambda x_{lp}}-e^{\lambda x_{lj}})(e^{\lambda u}+e^{\lambda
\bar{x}_{lj}})}}
\end{equation}

For real-valued $x_{pk}$ formula (\ref{Carleman}) becomes simpler:

\begin{equation}\label{RealCarleman}
\omega_m^\lambda(u,p,l) = 2 \frac{e^{\lambda u}-e^{\lambda x_{lp}}} {\lambda(e^{\lambda u}+e^{\lambda
x_{lp}})(u-x_{lp})}\times \prod_{j=1 j\neq p}^m{\frac{(e^{\lambda x_{lp}}+e^{\lambda
x_{lj}})(e^{\lambda u}-e^{\lambda x_{lj}})}{(e^{\lambda x_{lp}}-e^{\lambda x_{lj}})(e^{\lambda
u}+e^{\lambda x_{lj}})}}
\end{equation}

The Carleman's formula for extrapolation from $\mbox{\bf G}_M$ on $U=Q^n_\sigma$
($\sigma=\pi\lambda/2$) has the form ($z=(z_1,...,z_n)$):

\begin{equation}\label{Extr}
f_m(z) = \sum_{k_1,...,k_n=1}^m {f(x_k)\prod_{j=1}^n\omega^\lambda_m(z_j,k_j,j)},
\end{equation}

where $k={k_1,..,k_n}$, $x_k=(x_{1k_1},x_{2k_2},...,x_{nk_n})$.

There exists a theorem \cite{Aiz}:

{\bf If $f\in H^2(Q_\sigma^n)$, then $f(z) = lim_{m\rightarrow \infty} f_m(z)$, where
$H^2(Q_\sigma^n)$ is the Hardy class of holomorphic in $Q_\sigma^n$ functions. }

It is useful to present the asymptotics of (\ref{Extr}) for big $|\mbox{Re}z_j|$. For this we will
consider the asymptotics of (\ref{Extr}) for big $|\mbox{Re}u|$:

\begin{equation}\label{Asy}
|\omega_m^\lambda(u,p,l)| = \left|\frac{2}{\lambda u} \prod_{j=1 j\neq p}^m \frac{e^{\lambda
x_{lp}}+e^{\lambda x_{lj}}}{e^{\lambda x_{lp}}-e^{\lambda x_{lj}}}\right|+o(|\mbox{Re}u|^{-1}).
\end{equation}

From the formula (\ref{Extr}) one can see that for the finite $m$ and $|\mbox{Re}z_j|\rightarrow
\infty$ function $|f_m(z)|$ behaves like $const\cdot \prod_j |z_j|^{-1}$.

This property (null asymptotics) must be taken into account when using the formula (\ref{Extr}). When
constructing invariant manifolds $F(W)$, it is natural to use (\ref{Extr}) not for the immersion
$F(y)$, but for the deviation of $F(y)$ from some analytical ansatz  $F_0(y)$.

The analytical ansatz $F_0(y)$ can be obtained using Taylor series, just as in the Lyapunov auxiliary
theorem \cite{Lya}. Another variant is using Taylor series for the construction of
Pade-approximations.

It is natural to use approximations (\ref{Extr}) in dual variables as well, since there exists for
them (as the examples demonstrate) a simple and very effective linear ansatz for the invariant
manifold. This is the slow invariant subspace $E_{\mbox{\scriptsize slow}}$ of the operator of
linearized system (\ref{reaction}) in dual variables in the equilibrium point. This invariant
subspace corresponds to the the set of ``slow" eigenvalues (with small $|\mbox{Re}\lambda|$,
$\mbox{Re} \lambda<0$). In the initial space (of concentrations or densities) this invariant subspace
is the quasi-equilibrium manifold. It consist of the maximal entropy points on the affine manifolds
of the $x+E_{\mbox{\scriptsize fast}}$ form, where $E_{\mbox{\scriptsize fast}}$ is the ``fast"
invariant subspace of the operator of linearized system (\ref{reaction}) in the initial variables in
the equilibrium point. It corresponds to the ``fast" eigenvalues (big $|\mbox{Re}\lambda|$,
$\mbox{Re}\lambda<0$).

\subsection{Example: Two-step catalytic reaction}

Let's consider a two-step four-component reaction with one catalyst $A_2$:

\begin{equation}\label{tscatreact}
A_1+A_2 \leftrightarrow A_3 \leftrightarrow A_2+A_4
\end{equation}

We assume the Lyapunov function of the form $\mbox{\bf G}=\sum_{i=1}^4c_i[ln(c_i/c_i^*)-1]$. The
kinetic equation for the four-component vector of concentrations, ${\bf c}=(c_1,c_2,c_3,c_4)$, has
the form

\begin{equation}
\dot{{\bf c}} = \gamma_1W_1+\gamma_2W_2.
\end{equation}

Here $\gamma_{1,2}$ are stoichiometric vectors,

\begin{equation}
\gamma_1 = (-1,-1,1,0), \ \gamma_2=(0,1,-1,1),
\end{equation}

while functions $W_{1,2}$ are reaction rates:

\begin{equation}
W_1 = k_1^+c_1c_2-k_1^-c_3, \  W_2 = k_2^+c_3-k_2^-c_2c_4.
\end{equation}

Here $k_{1,2}^\pm$ are reaction rate constants. The system under consideration has two conservation
laws,

\begin{equation}
c_1+c_3+c_4 = B_1, \ c_2+c_3=B_2,
\end{equation}

or $\langle \bb_{1,2},\cc \rangle = B_{1,2}$, where $\bb_1 = (1,0,1,1)$ and $\bb_1 = (0,1,1,0)$. The
nonlinear system (\ref{tscatreact}) is effectively two-dimensional, and we consider a one-dimensional
reduced description. For our example, we chose the following set of parameters:

\begin{equation}
\begin{array}{lll}
k_1^+ = 0.3, \  k_1^- = 0.15, \ k_2^+ = 0.8, \ k_2^- = 2.0; \\ c_1^* = 0.5, \ c_2^* = 0.1, c_3^* =
0.1, \ c_4^* = 0.4;
\\ B_1 = 1.0, \ B_2 = 0.2
\end{array}
\end{equation}

In Fig.~\ref{4d1dgrid} one-dimensional invariant grid is shown in the ($c_1$,$c_4$,$c_3$)
coordinates. The grid was constructed by growing the grid, as described above. We used Newtonian
iterations to adjust the nodes. The grid was grown up to the boundaries of the phase space.

The grid derivatives for calculating tangent vectors $\bg$ were taken as simple as $\bg(\cc_i) =
(\cc_{i+1}-\cc_{i-1})/ \|\cc_{i+1}-\cc_{i-1}\|$ for the internal nodes and $\bg(\cc_1) =
(\cc_{1}-\cc_{2})/\| \cc_{1}-\cc_{2} \|$, $\bg(\cc_n) = (\cc_{n}-\cc_{n-1})/\| \cc_{n}-\cc_{n-1}\|$
for the grid's boundaries. Here $x_i$ denotes the vector of the $i$th node position, $n$ is the
number of nodes in the grid.

Close to the phase space boundaries we had to apply an adaptive algorithm for choosing the time step
$h$: if, after the next growing step and applying $N=20$ complete Newtonian steps, the grid did not
converged, then we choose a new $h_{n+1}=h_{n}/2$ and recalculate the grid. The final value for $h$
was $h \approx 0.001$.

The nodes positions are parametrized with entropic distance to the equilibrium point measured in the
quadratic metrics given by ${\HH_{\cc}} = ||\partial^2G({\cc})/\partial c_i\partial c_j||$ in the
equilibrium $\cc^*$. It means that every node is on a sphere in this quadratic metrics with a given
radius, which increases linearly. On this figure the step of the increase is chosen to be 0.05. Thus,
the first node is on the distance 0.05 from the equilibrium, the second is on the distance 0.10 and
so on. Fig.~\ref{4d1dgraphs} shows several basic values which facilitate understanding of the object
(invariant grid) extracted. The sign on the x-axis of the graphs at Fig.~\ref{4d1dgraphs} is
meaningless, since the distance is always positive, but in this situation it denotes two possible
directions from the equilibrium point.

Fig.~\ref{4d1dgraphs}a,b effectively represents the slow one-dimensional component of the dynamics of
the system. Given any initial condition, the system quickly finds the corresponding point on the
manifold and starting from this point the dynamics is given by a part of the graph on the
Fig.~\ref{4d1dgraphs}a,b.

One of the useful values is shown on the Fig.~\ref{4d1dgraphs}c. It is the relation between the
relaxation times ``toward" and ``along" the grid ($\lambda_2/\lambda_1$, where
$\lambda_1$,$\lambda_2$ are the smallest and the second smallest by absolute value non-zero
eigenvalue of the system, symmetrically linearized in the point of the grid node). It shows that the
system is very stiff close to the equilibrium point, and less stiff (by one order of magnitude) on
the borders. This leads to the conclusion that the reduced model is more adequate in the neighborhood
of the equilibrium where fast and slow motions are separated by two orders of magnitude. On the very
end of the grid which corresponds to the positive absciss values, our one-dimensional consideration
faces with definite problems (slow manifold is not well-defined).

\begin{figure}
\centering{
\caption{One-dimensional invariant grid (circles) for two-dimensional chemical system. Projection
into the 3d-space of $c_1$, $c_4$, $c_3$ concentrations. The trajectories of the system in the phase
space are shown by lines. The equilibrium point is marked by square. The system quickly reaches the
grid and further moves along it.} \label{4d1dgrid}}
\end{figure}

\begin{figure}
\centering{




\caption{One-dimensional invariant grid for two-dimensional chemical system. a) Values of the
concentrations along the grid. b) Values of the entropy ($-G$) and the entropy production ($-dG/dt$)
along the grid. c) Relation of the relaxation times ``toward" and ``along" the manifold. The nodes
positions are parametrized with entropic distance measured in the quadratic metrics given by ${\bf
H_c} = ||\partial^2G({\bf c})/\partial c_i\partial c_j||$ in the equilibrium $c^*$. Zero corresponds
to the equilibrium. }\label{4d1dgraphs}}
\end{figure}

\subsection{Example: Model hydrogen burning reaction}

In this section we consider a more interesting illustration, where the phase space is 6-dimensional,
and the system is 4-dimensional. We construct an invariant flag which consists of 1- and
2-dimensional invariant manifolds.

We consider chemical system with six species called (provisionally) $H_2$ (hydrogen), $O_2$ (oxygen),
$H_2O$ (water), $H$, $O$, $OH$ (radicals). We assume the Lyapunov function of the form
$G=\sum_{i=1}^6c_i[ln(c_i/c_i^*)-1]$. The subset of the hydrogen burning reaction and corresponding
(direct) rate constants have been taken as:

\begin{equation}
\begin{array}{llllll}
1.\hspace{0.1cm} H_2 \leftrightarrow 2H \hspace{0.5cm} & k_1^+ = 2
\\

2.\hspace{0.1cm} O_2 \leftrightarrow 2O \hspace{0.5cm} & k_2^+ = 1
\\

3.\hspace{0.1cm} H_2O \leftrightarrow H+OH \hspace{0.5cm} & k_3^+ = 1 \\

4.\hspace{0.1cm} H_2+O \leftrightarrow H+OH \hspace{0.5cm} & k_4^+ = 10^3 \\

5.\hspace{0.1cm} O_2+H \leftrightarrow O+OH \hspace{0.5cm} & k_5^+ = 10^3 \\

6.\hspace{0.1cm} H_2+O \leftrightarrow H_2O \hspace{0.5cm} & k_6^+ = 10^2
\end{array}
\end{equation}

The conservation laws are:

\begin{equation}
\begin{array}{ll}
2c_{H_2}+2c_{H_2O}+c_H+c_OH = b_H \\

2c_{O_2}+c_{H2O}+c_O+c_OH = b_O
\end{array}
\end{equation}

For parameter values we took  $b_H=2$, $b_O=1$, and the equilibrium point:

\begin{equation}
\begin{array}{llllll}
c_{H_2}^*=0.27 & c_{O_2}^* = 0.135 & c_{H_2O}^*=0.7 & c_H^*=0.05 & c_O^*=0.02 & c_{OH}^*=0.01
\end{array}
\end{equation}

Other rate constants $k_i^{-},i=1..6$ were calculated from $\cc^*$ value and $k_i^{+}$. For this
system the stoichiometric vectors are:

\begin{equation}
\begin{array}{llllll}
\gamma_1 = (-1,0,0,2,0,0) & \gamma_2 = (0,-1,0,0,2,0) \\

\gamma_3 = (0,0,-1,1,0,1) & \gamma_4 = (-1,0,0,1,-1,1) \\

\gamma_5 = (0,-1,0,-1,1,1) & \gamma_6 = (-1,0,1,0,-1,0)
\end{array}
\end{equation}

We stress here once again that the system under consideration is fictional in that sense that the
subset of equations corresponds to the simplified picture of this physical-chemical process and the
constants do not correspond to any measured ones, but reflect only basic orders of magnitudes of the
real-world system. In this sense we consider here a qualitative model system, which allows us to
illustrate the invariant grids method without excessive complication. Nevertheless, modeling of real
systems differs only in the number of species and equations. This leads, of course, to
computationally harder problems, but not the crucial ones, and the efforts on the modeling of
real-world systems are on the way.

Fig.~\ref{6d1dgrid}a presents a one-dimensional invariant grid constructed for the system.
Fig.~\ref{6d1dgrid}b shows the picture of reduced dynamics along the manifold (for the explanation of
the meaning of the $x$-coordinate, see the previous subsection). On Fig.~\ref{6d1dgrid}c the three
smallest by absolute value non-zero eigen values of the symmetrically linearized system $A^{sym}$
have been shown. One can see that the two smallest values ``exchange" on one of the grid end. It
means that one-dimensional "slow" manifold has definite problems in this region, it is just not
defined there. In practice, it means that one has to use at least two-dimensional grids there.

\begin{figure}
\centering{




\caption{One-dimensional invariant grid for model hydrogen burning system. a) Projection into the
3d-space of $c_H$, $c_O$, $c_{OH}$ concentrations. b) Concentration values along the grid. c) three
smallest by absolute value non-zero eigen values of the symmetrically linearized system.
}\label{6d1dgrid}}
\end{figure}

Fig.~\ref{6d2dgrid}a gives a view onto the two-dimensional invariant grid, constructed for the
system, using the ``invariant flag" strategy. The grid was grown starting from the 1D-grid
constructed at the previous step. At the first iteration for every node of the initial grid, two
nodes (and two edges) were added. The direction of the step was chosen as the direction of the
eigenvector of the matrix $A^{sym}$ (in the point of the node), corresponding to the second
``slowest" direction. The value of the step was chosen to be $\epsilon=0.05$ in terms of entropic
distance. After several Newtonian iterations done until convergence, new nodes were added in the
direction ``ortogonal" to the 1D-grid. This time it is done by linear extrapolation of the grid on
the same step $\epsilon=0.05$. When some new nodes have one or several negative coordinates (the grid
reaches the boundaries) they were cut off. If a new node has only one edge, connecting it to the
grid, it was excluded (since it does not allow calculating 2D-tangent space for this node). The
process continues until the expansion is possible (after this, every new node has to be cut off).

\begin{figure}
\centering{



\caption{Two-dimensional invariant grid for the model hydrogen burning system. a) Projection into the
3d-space of $c_H$, $c_O$, $c_{OH}$ concentrations. b) Projection into the principal 3D-subspace.
Trajectories of the system are shown coming out from the every grid node. Bold line denotes the
one-dimensional invariant grid, starting from which the 2D-grid was constructed.} \label{6d2dgrid}}
\end{figure}

Strategy of calculating tangent vectors for this regular rectangular 2D-grid was chosen to be quite
simple. The grid consists of {\it rows}, which are co-oriented by construction to the initial
1D-grid, and {\it columns} that consist of the adjacent nodes in the neighboring rows. The direction
of ``columns" corresponds to the second slowest direction along the grid. Then, every row and column
is considered as 1D-grid, and the corresponding tangent vectors are calculated as it was described
before: $$\bg_{row}(\cc_{k,i}) = (\cc_{k,i+1}-\cc_{k,i-1})/\| \cc_{k,i+1}-\cc_{k,i-1} \|$$ for the
internal nodes and $$\bg_{row}(\cc_{k,1}) = (\cc_{k,1}-\cc_{k,2})/\| \cc_{k,1}-\cc_{k,2}\|,
\bg_{row}(\cc_{k,n_k}) = (\cc_{k,n_k}-\cc_{k,n_k-1})/\| \cc_{k,n_k}-\cc_{k,n_k-1} \|$$ for the nodes
which are close to the grid's edges. Here $x_{k,i}$ denotes the vector of the node in the $k$th row,
$i$th column; $n_k$ is the number of nodes in the $k$th row. Second tangent vector
$\bg_{col}(\cc_{k,i})$ is calculated completely analogously. In practice, it is convenient to
orthogonalize $\bg_{row}(\cc_{k,i})$ and $g_{col}(\cc_{k,i})$.

Since the phase space is four-dimensional, it is impossible to visualize the grid in one of the
coordinate 3D-views, as it was done in the previous subsection. To facilitate visualization one can
utilize traditional methods of multi-dimensional data visualization. Here we make use of the
principal components analysis (see, for example, \cite{princ}), which constructs a three-dimensional
linear subspace with maximal dispersion of the othogonally projected data (grid nodes in our case).
In other words, method of principal components constructs in multi-dimensional space such a
three-dimensional box inside which the grid can be placed maximally tightly (in the mean square
distance meaning). After projection of the grid nodes into this space, we get more or less adequate
representation of the two-dimensional grid embedded into the six-dimensional concentrations space
(Fig.~\ref{6d2dgrid}b). The disadvantage of the approach is that the axes now do not have explicit
meaning, being some linear combinations of the concentrations.

One attractive feature of two-dimensional grids is the possibility to use them as a screen, on which
one can display different functions $f(\cc)$ defined in the concentrations space. This technology was
exploited widely in the non-linear data analysis by the elastic maps method \cite{GZ01}. The idea is
to ``unfold" the grid on a plane (to present it in the two-dimensional space, where the nodes form a
regular lattice). In other words, we are going to work in the internal coordinates of the grid. In
our case, the first internal coordinate (let's call it $s_1$) corresponds to the direction,
co-oriented with the one-dimensional invariant grid, the second one (let's call it $s_2$) corresponds
to the second slow direction. By how it was constructed, $s_2=0$ line corresponds to the
one-dimensional invariant grid. Units of $s_1$ and $s_2$ are entropic distances in our case.

Every grid node has two internal coordinates $(s_1,s_2)$ and, simultaneously, corresponds to a vector
in the concentration space. This allows us to map any function $f({\bf c})$ from the
multi-dimensional concentration space to the two-dimensional space of the grid. This mapping is
defined in a finite number of points (grid nodes), and can be interpolated (linearly, in the simplest
case) in between them. Using coloring and isolines one can visualize the values of the function in
the neighborhood of the invariant manifold. This is meaningful, since, by the definition, the system
spends most of the time in the vicinity of the invariant manifold, thus, one can visualize the
behaviour of the system. As a result of applying the technology, one obtains a set of color
illustrations (a stack of information layers), put onto the grid as a map. This allows applying all
the methods, working with stack of information layers, like geographical information systems (GIS)
methods, which are very well developed.

In short words, the technique is a useful tool for exploration of dynamical systems. It allows to see
simultaneously many different scenarios of the system behaviour, together with different system's
characteristics.

The simplest functions to visualize are the coordinates: $c_i(\cc) = c_i$. On
Fig.~\ref{6d2dgridcolor} we displayed four colorings, corresponding to the four arbitrarily chosen
concentrations functions (of $H_2$, $O$, $H$ and $OH$; Fig.~\ref{6d2dgridcolor}a-d). The qualitative
conclusions that can be made from the graphs are that, for example, the concentration of $H_2$
practically does not change during the first fast motion (towards the 1D-grid) and then, gradually
changes to the equilibrium value (the $H_2$ coordinate is ``slow"). The $O$ coordinate is the
opposite case, it is ``fast" coordinate which changes quickly (on the first stage of motion) to the
almost equilibrium value, and then it almost does not change. Basically, the slope angles of the
coordinate isolines give some presentation of how ``slow" a given concentration is.
Fig.~\ref{6d2dgridcolor}c shows interesting behaviour of the $OH$ concentration. Close to the 1D grid
it behaves like ``slow coordinate", but there is a region on the map where it has clear ``fast"
behaviour (middle bottom of the graph).

The next two functions which one can want to visualize are the entropy $S=-G$ and the entropy
production $\sigma(\cc)=-dG/dt(\cc) = - \sum_i{\ln(c_i/c^*_i){\dot c_i}}$. They are shown on
Fig.~\ref{6d2dgridcolor1}a,b.

Finally, we visualize the relation between the relaxation times of the fast motion towards the
2D-grid and along it. This is given on the Fig.~\ref{6d2dgridcolor1}c. This picture allows to make a
conclusion that two-dimensional consideration can be appropriate for the system (especially in the
``high $H_2$, high $O$" region), since the relaxation times ``towards" and ``along" the grid are
definitely separated. One can compare this to the Fig.~\ref{6d2dgridcolor1}d, where the relation
between relaxation times towards and along the 1D-grid is shown.

\begin{figure}
\centering{

a) Concentration $H_2$ \hspace{2.5cm} b) Concentration $O$ \\
c) Concentration $OH$ \hspace{2.5cm} d) Concentration $H$

\caption{Two-dimensional invariant grid as a screen for visualizing different functions defined in
the concentrations space.
The coordinate axes are entropic distances (see the text for the explanations) along the first and
the second slowest directions on the grid. The corresponding 1D invariant grid is denoted by bold
line, the equilibrium is denoted by square. }\label{6d2dgridcolor}}
\end{figure}

\begin{figure}
\centering{

a) Entropy \hspace{3.5cm} b) Entropy Production \\
c) $\lambda_3/\lambda_2$ relation \hspace{3.5cm} d) $\lambda_2/\lambda_1$ relation

\caption{Two-dimensional invariant grid as a screen for visualizing different functions defined in
the concentrations space.
The coordinate axes are entropic distances (see the text for the explanations) along the first and
the second slowest directions on the grid. The corresponding 1D invariant grid is denoted by bold
line, the equilibrium is denoted by square.}\label{6d2dgridcolor1}}
\end{figure}

\section{Method of invariant manifold for open systems}
\label{open}

One of the problems to be focused on when studying closed systems is to prepare extensions of the
result for open or driven by flows systems. External flows are usually taken into account by
additional terms in the kinetic equations (\ref{reaction}):
\begin{equation}
\label{external} \dot{\cc}=\JJ(\cc)+\bPi.
\end{equation}
{\it Zero-order approximation} assumes that the flow does not change the invariant manifold.
Equations of the reduced dynamics, however, do change: Instead of $\JJ(\cc(M))$ we substitute
$\JJ(\cc(M))+\bPi$ into Eq.\ (\ref{dyn}):
\begin{equation}
\label{zero} \dot{M}_i=(\bnabla M_i\big|_{\cc(M)}, \JJ(\cc(M))+\bPi).
\end{equation}
Zero-order approximation assumes that the fast dynamics in the closed system strongly couples the
variables $\cc$, so that flows cannot influence this coupling.

{\it First-order approximation} takes into account the shift of the invariant manifold by
$\delta\cc$. Equations for Newton's iterations have the same form (\ref{ITERATION}) but instead of
the vector field $\JJ$ they take into account the presence of the flow:
\begin{equation}
\label{one} [1-\PP_{\cc}](\bPi+\LL'_{\cc}\delta\cc)=0,\ \PP_{\cc}\delta\cc=0,
\end{equation}
where projector $\PP_{\cc}$ corresponds to the unperturbed manifold.

The first-order approximation means that fluxes change the coupling between the variables
(concentrations). It is assumed that these new coupling is also set instantaneously (neglect of
inertia).

{\it Remark.} Various realizations of the first-order approximation in physical and chemical dynamics
implement the viewpoint of an infinitely small chemical reactor driven by the flow. In other words,
this approximation is applicable in the Lagrangian system of coordinates \cite{KGDN98,ZKD00}.
Transition to Eulerian coordinates is possible but the relations between concentrations and the flow
will change its form. In a contrast, the more simplistic zero-order approximation is equally
applicable in both the coordinate system, if it is valid.

\section{Conclusion}
\label{conclusion}

In this paper, we have presented the method for constructing the invariant manifolds for reducing
systems of chemical kinetics. Our approach to computations of invariant manifolds of dissipative
systems is close in spirit to the Kolmogorov-Arnold-Moser theory of invariant tori of Hamiltonian
systems \cite{Arnold63,Arnold83}: We also base our consideration on the Newton method instead of
Taylor series expansions \cite{Beyn98}, and systematically use duality structures. Recently, a
version of an approach based on the invariance equations was rediscovered in \cite{Kazantzis00}. He
was solving  the invariance equation by a Taylor series expansion. A counterpart of Taylor series
expansions for constructing the slow invariant manifolds in the classical kinetic theory is the
famous Chapman-Enskog method. The question of how this compares to iteration methods was studied
extensively for certain classes of Grad moment equations \cite{GK96a,KDN97a,K00}.

The thermodynamic parameterization and the selfadjoint linearization arise in a natural way in the
problem of finding slowest invariant manifolds for closed systems. This also leads to  various
applications in different approaches to reducing the description, in particular, to a
thermodynamically consistent version of the intrinsic low-dimensional manifold, and to  model kinetic
equations for lifting the reduced dynamics. Use of the thermodynamic projector makes it unnecessary
global parameterizations of manifolds, and thus leads to computationally promising grid-based
realizations.

Invariant manifolds are constructed for closed space-independent chemical systems. We also describe
how to use these manifolds for modeling open and distributed systems.

{\bf Acknowledgements.} We thank Dr. Vladimir Zmievski (Ecole Polytechnique Montreal) for
computations in section \ref{EX}. Fruitful discussions with Prof.\ Hans Christian \"Ottinger (ETH
Zurich) on the problem of reduction are gratefully acknowledged. Encouragement of Prof.\ Misha Gromov
(IHES Bures-sur-Yvette) was very important for completing this work.

\end{document}